\documentclass[journal=jacsat,manuscript=article]{achemso}
\usepackage{graphicx}

\usepackage{dcolumn}
\usepackage{bm}
\usepackage{caption}
\usepackage{subcaption}
\graphicspath{{figures/}} 

\title{A simple approach to hopping matrix elements between nearby molecules} 
\author{Shane Donaher, Puja Agarwala, Scott T. Milner}
\affiliation{Department of Chemical Engineering, The Pennsylvania State University, 
University Park, PA, 16802}

\begin{document}

\begin{abstract}
The hopping matrix element $t$ quantifies the quantum-mechanical coupling
between frontier orbitals on a pair of nearby molecules.
Convenient and generally applicable methods 
to determine $t$ from DFT calculations are lacking;
$t$ can be obtained from coupling-induced energy splittings
only if the interacting molecules are identical and symmetrically placed.
We present a simple approach to determine $t$ from DFT results
that relies on measuring hybridization,
projecting hybridized pair orbitals 
onto constituent frontier orbitals of the interacting molecules,
using spatially discretized wavefunctions (``cube files'')
rather than analytical representations.
We demonstrate the method by exploring how $t$ depends
on the identity and relative placement 
of typical moieties found in semiconducting polymers.
\end{abstract}

\section{Introduction}
The hopping matrix element $t$ quantifies the quantum-mechanical coupling
between frontier orbitals on a pair of nearby molecules.
$t$ is important in describing transfer of electrons or holes
between moieties in close proximity,
which governs the mobility of carriers in molecular solids\cite{Kwon:tt}.
In ordered molecular solids, $t$ controls the valence and conduction bandwidths\cite{Kwon:tt},
and thereby determines the group velocity of electron and hole wave packets\cite{Cheng:2003bs}.
In disordered materials, $t$ influences the diffusivity of holes and electrons, 
which in turn governs their drift mobilities\cite{Deng:ud}.

Because $t$ depends sensitively on geometry and molecular structure,
it would be useful to have a simple, efficient, generally applicable method 
to compute $t$ between arbitrary pairs of small molecules in arbitrary geometries.
Such a method could for example be applied 
to explore heterogeniety in disordered molecular solids,
by computing $t$ for many thousands of pairs of moieties in different arrangements, 
extracted from configurations obtained from molecular dynamics simulations.
Extensive sampling of $t$ over many arrangements could be useful
to investigate heterogeniety in charge transport,
or in the energetics of charge-transfer excitons at bulk heterojunction interfaces,
or to gain insight as to what molecular structures and arrangements give large values of $t$\cite{Troisi:us,miller2019machine,wang2020artificial}.

Indeed, a common task in quantum chemistry calculations more generally
is to determine the Hamiltonian matrix element
that couples diabatic states on two nearby moieties.
(By ``diabatic states'', here we mean frontier orbitals on each of the two moieties,
which are eigenstates for each moiety in isolation, but not for the interacting pair.)

Because of their central importance,
many techniques have been developed to calculate hopping matrix elements.
One longstanding approach is the Mulliken-Hush method,
originating in discussions of electron transfer reactions
\cite{Hush1968,Mulliken1952,Cave1996,Cave1997}.
(A useful pedagogical treatment of the MH method 
is given by Creutz et al.\cite{Creutz1994}.)
The essence of the method is as follows.
First, a simplified 2x2 matrix Hamiltonian is written
to represent an interacting system 
of localized states on each of two nearby moieties $a$ and $b$,
for which bonding and antibonding eigenstates and energies 
can be computed exactly in terms of the matrix elements.

Then, the dipole matrix element coupling the bonding and antibonding states
is related to the Hamiltonian matrix element between the two localized states,
under two physically reasonable assumptions: 
1) weak direct overlap between the localized states,
so that $\mu_{ab}$ is small 
compared to the separation between the two moieties;
and 2) the dipole matrix element between the bonding and antibonding states
points along the separation vector between the two moieties.

This argument results in the relation
\begin{equation}
H_{ab} = \frac{ \mu_{12} \Delta E_{12}}{\Delta \mu_{ab}} 
= \frac{\mu_{12} \Delta E_{12}}{(\Delta \mu_{12}^2 + 4 \mu_{12}^2 )^{1/2}}
\end{equation}
Here subscripts $a$ and $b$ refer to the moieties,
and 1 and 2 refer to the bonding and antibonding states
of the interacting pair;
$\Delta E_{12} = E_1 - E_2$ is the splitting;
and $\mu_{12}$ the dipole matrix element coupling the bonding and antibonding states.
Finally, $\mu_{12} = \mu_1 - \mu_2$ is the difference in dipole moments 
of the two diabatic states, measured with respect to a common origin,
and typically approximated by the separation vector between the moieties
\cite{Cave1996,Cave1997}.
This expression was originally used to obtain 
Hamiltonian matrix elements from experimental absorption intensities,
but has also been applied to compute Hamiltonian matrix elements 
from dipole matrix elements and energy differences
computed using density functional theory.

Following the work of Mulliken and Hush,
many extensions, improvements, and new approaches
have been introduced for computing matrix elements 
between nearby moieties.

The MH method has been generalized by Cave and Newton
to systems in which more than two localized states interact
\cite{Cave1996,Cave1997}.
Voityuk et al.\ introduced an analysis analogous to MH,
but focusing on charge transfer between moieties
rather than dipole matrix elements,
to relate the Hamiltonian matrix element to orbital energy differences
and charge transfer between donor and acceptor 
resulting from an excitation 
from the bonding to the antibonding eigenstate
\cite{Voityuk2002}.
With the advent of widespread density functional codes,
direct evaluation of Hamiltonian matrix elements became practical,
by analytically computing matrix elements of the Fock operator
(i.e., the effective 1-body Hamiltonian, obtained from an SCFT treatment).
Related approaches focus on more efficient analytical 
evaluations of matrix elements the effective 1-body Hamiltonian
for some representation of the interacting states,
for example the fragment-orbital DFT approach of Senthilkumar et al.
\cite{Senthilkumar2003}.
However, these approaches are not readily available 
to end users of commonly available DFT platforms such as Gaussian,
which do not give easy access to Fock operator matrix elements
between frontier orbitals.

Kirkpatrick developed an approximate method using ZINDO,
which avoids a full SCF treatment of the pair of interacting moieties,
requiring only an SCF treatment of each moiety separately
\cite{Kirkpatrick2008}.
However, this approach is limited by the ZINDO semi-empirical approximation,
which results in factor-of-two discrepancies with more exact calculations,
while properly capturing trends with changing molecular arrangements.

Finally, in some cases it can be challenging
to obtain appropriate diabatic states;
Wu and van Voorhis address this problem
by applying constrained DFT to force charges to reside 
exclusively on the starting and ending moieties
\cite{Wu2006,Wu2006b}.

In this paper, we present a simple numerical approach to extract $t$ 
from DFT calculations of orbital energies and wavefunctions.
The method uses results generated by the quantum chemistry package Gaussian,
and requires no sophisticated coding or access to orbital expansion coefficients.

Compared to previous methods briefly discussed above, 
our elementary approach relates most closely to the original MH method.
However, we circumvent the need for dipole matrix elements
(and attendant approximations of MH),
preferring to work directly with the effective 2x2 system,
as characterized by readily accessible DFT results,
namely the orbital energies and wavefunction cube files.

In essence, we project the complex multielectron system
to the 2x2 space of interacting frontier orbitals on each moiety,
both with respect to the Hamiltonian
and the resulting frontier orbitals of the interacting pair,
consistent with the spirit of tight-binding models.
It can be shown that our approach reduces to MH, 
in limit of localized diabatic states well separated in space
and transition dipole pointed along the separation vector.

To demonstrate the utility of our method, we use it to investigate how $t$ depends 
on the relative placement and identity of nearby pairs of molecules. 
Previous investigations of this sort have been performed 
on pairs of identical molecules in idealized geometries,
including a pair of ethylene molecules
\cite{Valeev2006,Kirkpatrick2008},
pairs of parallel-oriented pentacene or hexabenzocoronene molecules
\cite{Kirkpatrick2008},
and neighboring oligoacenes in a crystal
.\cite{JeanLucBredas:2004bt,Kwon:tt,AlessandroTroisi:2005et,
Troisi:us,Hummer:2005kr,Tiago:2003dm}.

In what follows, we first validate our approach
by comparing to previous results for tetracene and ethylene.
Then, we investigate a variety of molecular pairs, 
both identical and dissimilar,
with structures relevant to organic photovoltaics.
For a given molecular pair, we perform scans 
in which we either translate or rotate one molecule with respect to the other.
In such scans, the onsite energy of the molecules in the pair
and the relative placement of their hybridizing orbitals vary systematically,
both of which in turn affect $t$.
From such scans, we can gain insight 
as to what electronic and geometric factors
lead to sizable hopping matrix elements between nearby molecules. 

\section{Orbital interactions}

When two molecules are brought close together,
their frontier orbitals hybridize to form the frontier orbitals of the pair.
Typically, the frontier orbitals of the pair consist predominantly 
of one particular orbital from each molecule;
for example, the HOMO of a pair may consist mainly 
of a linear combination of the HOMOs on each molecule. 
Our calculations and discussion make this simplifying assumption,
which can be checked as part of the calculation. 

Quantum packages and orbital visualizations give insight 
into how individual molecule orbitals hybridize to form pair orbitals.
Pair orbital images can typically be recognized
as made from identifiable orbitals of the two individual molecules,
and clearly display how the wavefunction is distributed between the two.  
In this paper, the two molecules within a pair will be referred to
as the ``monomers'' that constitute the pair.

\begin{figure}[htbp]
\begin{center}
\includegraphics[width=0.45\textwidth]{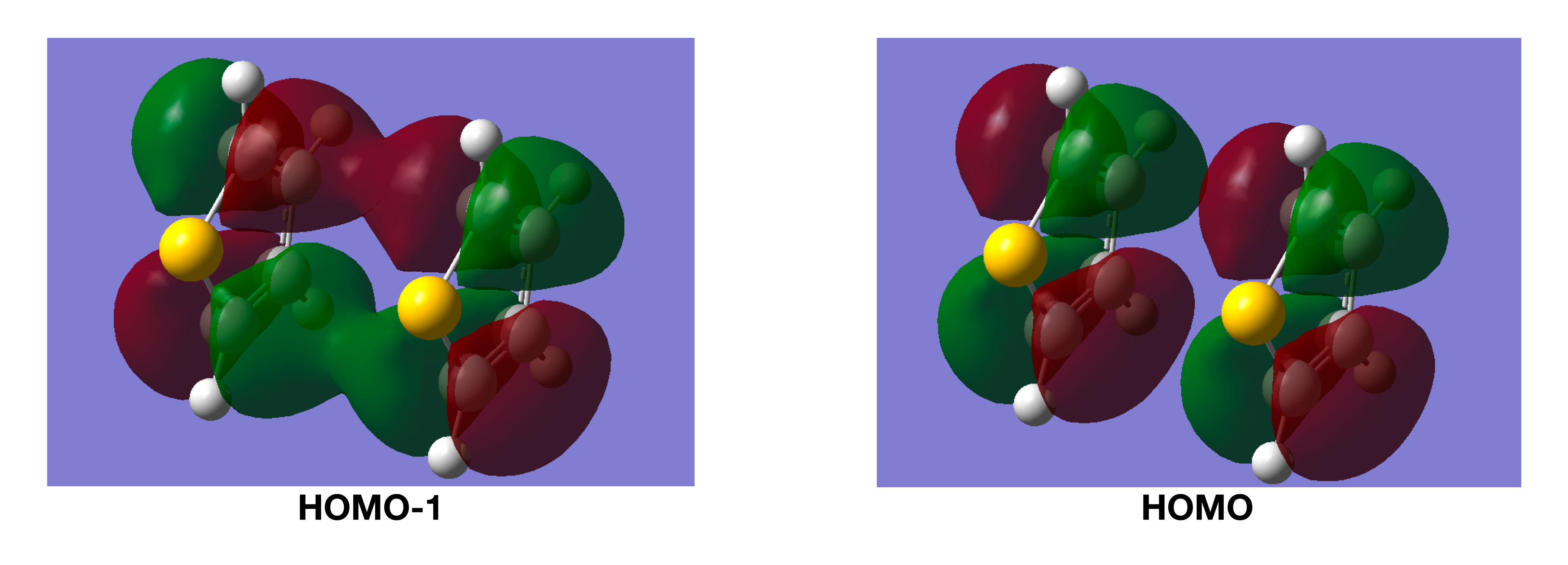}
\caption{HOMO-1 and HOMO of thiophene-thiophene pair at 4.0 \AA\ separation.}
\label{fig:ThioBondAntiEx}
\end{center}
\end{figure}

Fig.\ \ref{fig:ThioBondAntiEx}
shows how the HOMOs of two thiophene monomers hybridize
to form the HOMO and HOMO-1 of the pair. 
The pair HOMO-1 in Fig.\ \ref{fig:ThioBondAntiEx}
is evidently a bonding orbital of the molecular HOMOs,
in which the lobes of the wavefunction on each molecule are adjacent
to lobes of the same sign on the opposing molecule.
This arrangement avoids gradients within the pair orbital wavefunction,
lowering its kinetic energy.
By ``lobe'', we mean a region of the wavefunction of a given sign,
indicated by color in Fig.\ \ref{fig:ThioBondAntiEx}.
The thiophene HOMO has four lobes. 

Likewise, in the HOMO of Fig.\ \ref{fig:ThioBondAntiEx},
the lobes on one molecule are adjacent to lobes of opposite sign on the opposing molecule,
forming an antibonding orbital.
This arrangement gives rise to nodes between opposing lobes,
resulting in wavefunction gradients that raise the kinetic energy. 

The example of Fig.\ \ref{fig:ThioBondAntiEx} is a special case,
in which the constituent monomers are identical and symmetrically related.
As a consequence, the pair HOMO and HOMO-1 wavefunctions 
have mirror symmetry with respect to the two molecules
and thus equal magnitude on both molecules,
as evident in the figure.
This mirror symmetry holds regardless of how strongly or weakly 
the orbitals interact with one another through hybridization. 

For this special case,
the onsite energies of the constituent monomers must be equal by symmetry,
and $t$ can be calculated directly from the splitting 
of the bonding and antibonding pair orbitals
\cite{Valeev2006}.
This symmetry holds for identical molecules,
so long as they are symmetrically placed. 
For example, if we translate or rotate the molecules in Fig.\ \ref{fig:ThioBondAntiEx}
with respect to the central orthogonal axis,
the arrangement remains symmetric.

However, not all arrangements of identical molecules are symmetric.
In general, each monomer perturbs the orbital energies of the other monomer to some extent.
When the monomers are not symmetrically placed,
these perturbations need not be equal,
and so the effective values of the onsite energy on each monomer may differ.
As a consequence, the pair orbitals will no longer have mirror symmetry,
and the wavefunction may be unequally distributed between the two monomers.

\begin{figure}[htbp]
\begin{center}
\includegraphics[width=0.45\textwidth]{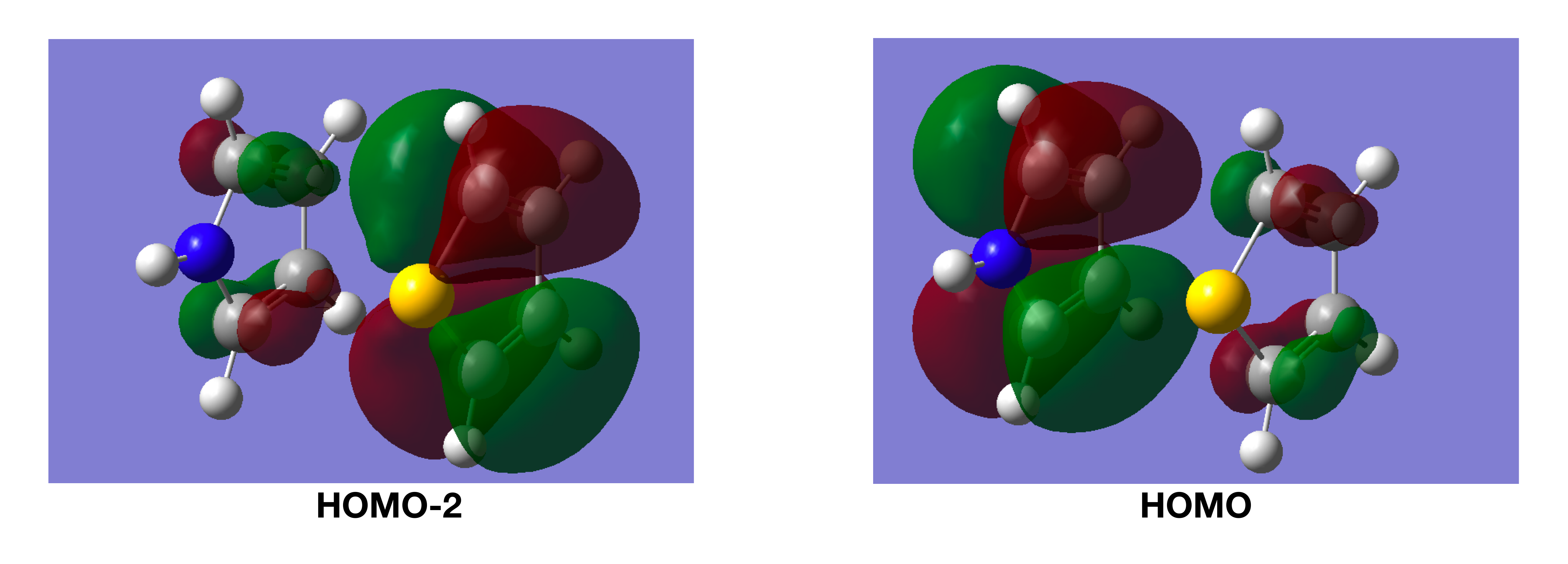}
\caption{HOMO-2 and HOMO of thiophene-pyrrole pair at 4.0 \AA\ separation.}
\label{fig:ThPyrOrbitals}
\end{center}
\end{figure}

Of course, asymmetric distributions of pair orbitals also result from using different constituent monomers,
as shown for a pyrrole-thiophene pair in Fig.\ \ref{fig:ThPyrOrbitals}.
Once again, the HOMO of thiophene hybridizes with the HOMO of pyrrole within the pair.
However, now the resulting bonding and antibonding orbitals form the HOMO-2 and the HOMO of the pair
(the pair HOMO-1 is an unrelated orbital, not participating in this hybridization).
The pyrrole HOMO is significantly higher in energy than for thiophene,
but the orbitals have roughly the same shape. 

In Fig.\ \ref{fig:ThPyrOrbitals},
the pair orbital wavefunction is not equally distributed between the two moieties.
The lower-energy bonding orbital resides mainly on the lower-energy HOMO of thiophene,
while the antibonding orbital resides primarily on the higher-energy HOMO of pyrrole.
Importantly, each monomer perturbs the onsite energy of the other,
to an unknown and unequal extent.
Thus, we cannot assume the onsite energies of the two molecules in proximity
are the same as when isolated,
so we cannot calculate $t$ from the pair orbital energy splitting alone.
To compute $t$ in such cases,
we need an additional piece of information,
as we shall show in the next section. 

\section{Tight-binding model}

To represent the hybridization of one orbital on each of two molecules,
we write a simple $2 \times 2$ tight-binding model,
for which the Hamiltonian matrix is
\begin{equation}
H = \left\lbrack
\begin{array}{ccc}
  \epsilon_1& -$t$ \\
  -$t$& \epsilon_2  \\
\end{array}
\right\rbrack
\label{eqn:Hamiltonian Matrix Eq}
\end{equation}
Here $ \epsilon_1 $ and $ \epsilon_2 $ are the onsite energies 
of the relevant frontier orbitals on the two monomers,
and $t$ the hopping matrix element between these orbitals.
Our sign convention is that $t>0$ corresponds to the case
in which monomer orbitals with amplitudes of the same sign form a bonding orbital. 

The simple two-state Hamiltonian in Eqn.\ \ref{eqn:Hamiltonian Matrix Eq}
is evidently a drastic simplification of the full quantum mechanics
describing the pair of interacting molecules;
the active space is projected down to a single pair of orbitals.
The frontier orbital of the pair is correspondingly written 
as a linear combination of the two interacting monomer orbitals.
Nonetheless, this simple two-site model reasonably represents 
interacting molecules that are close enough to hybridize
but not so close that their charge distributions drastically overlap.  

The energy eigenvalues of the 2x2 Hamiltonian take the familiar form
\begin{equation}
	\lambda_\pm=\frac{\epsilon_1 + \epsilon_2 \pm \sqrt{(\epsilon_1 - \epsilon_2)^2+4t^2}}{2}
	\label{eqn:EigenvaluesHam}
\end{equation}
with the splitting between the two states $ \Delta E = \lambda_{+} - \lambda_{-} $ given as
\begin{equation}
	\Delta E=\sqrt{(\epsilon_1-\epsilon_2)^2+4t^2}
	\label{eqn:EigenvaluesDiff}
\end{equation}

Evidently, the splitting $\Delta E$ depends on both $t$ 
and the onsite energy difference $\epsilon_1 - \epsilon_2$.
Only for the special case of identical monomers symmetrically arranged 
does $\epsilon_1 - \epsilon_2$ vanish,
whereupon the splitting alone determines $t$, as $|t|= \Delta E/2$.
This limit corresponds to the energy-splitting in dimer (ESD) method
\cite{ottonelli2012koopmans}.
Otherwise, when the onsite energies are not forced by symmetry to be equivalent,
we need more information about the pair system to calculate $t$.
In short, we get this information by measuring the coefficients of the pair wavefunction.  

The corresponding eigenvectors $e^+ = \{e^+_1, e^+_2\}$ and $e^- = \{e^-_1, e^-_2\}$ 
of $H$ take the form
\begin{equation}
\left\lbrack
\begin{array}{ccc}
  e^+_1& e^-_1 \\
  e^+_2&e^-_2  \\
\end{array}
\right\rbrack = \frac{1}{\sqrt{1+\alpha^2}}\left\lbrack
\begin{array}{ccc}
  1& \alpha \\
  -\alpha& 1  \\
\end{array}
\right\rbrack
\label{eqn:HamiltonianEigenvectors}
\end{equation}
in which $\alpha$ is given by
 \begin{equation}
	\alpha=\frac{1}{2t}(\sqrt{(\epsilon_1-\epsilon_2)^2+4t^2} - (\epsilon_1-\epsilon_2))
	\label{eqn:Diffebefsdaa}
\end{equation}

Here we have adopted conventions such that $\epsilon_1 > \epsilon_2$,
and interpret the square root as positive, 
so that $0 \leq \alpha \leq 1$,
and $\lambda_+$ approaches $\epsilon_1$ as $t$ becomes small.
Solving Eqn.\ \ref{eqn:EigenvaluesDiff} and Eqn.\ \ref{eqn:Diffebefsdaa} 
for the onsite energy difference and $t$ in terms of $\alpha$ and $\Delta E$,
we obtain
 \begin{equation}
	\epsilon_1-\epsilon_2=\frac{1-\alpha^2}{1+\alpha^2}\Delta E
	\label{eqn:Diffebea}
\end{equation}
\begin{equation}
	t=\frac{\alpha}{1+\alpha^2}\Delta E
	\label{eqn:Finalt}
\end{equation}

We extract $\alpha$ from DFT results 
by projecting the relevant pair orbitals onto the constituent monomer orbitals
to obtain the $2 \times 2$ matrix of eigenvectors.
Thus the eigenvector coefficients $e^\pm_i$ in Eqn.\ \ref{eqn:HamiltonianEigenvectors}
are given by
\begin{equation}
e^\pm_i = \langle \psi_\pm | \phi_i \rangle
\label{eq:project}
\end{equation}
in which $\psi_+$ and $\psi_-$ are the two hybridized pair orbitals,
and $\phi_1$ and $\phi_2$ the constituent orbitals on the two monomers.

In writing Eqn.\ \ref{eq:project}, we are assuming for simplicity
that the frontier orbitals on the two monomers have negligible overlap,
compared to their projections onto the hybridized dimer orbital.
This is the case for all examples presented in this paper.
More generally, we can construct using standard linear algebra techniques
a dual basis of two mostly-localized states $\{ \tilde \psi_i \}$
that satisfy $\langle \tilde \psi_i | \psi_j \rangle = \delta_{ij}$,
which we can use to extract the expansion coefficients $\{ e^{\pm}_i \}$.
See Appendix for details.

We calculate the projections numerically,
representing the orbitals to be integrated 
as spatially discretized ``cube files'',
which can be readily obtained from Gaussian. 
To facilitate this approach,
we ensure that the cube files for the pair and both monomer orbitals 
are computed on the same grid of sampling points.
In our Gaussian calculations,
we use the B3LYP function with the 6-31g(d,p) basis set. 
See Supplemental Information for details.

If our simple $2 \times 2$ model adequately represents the real system,
the eigenvectors generated should be approximately in the form
given by Eqn.\ \ref{eqn:HamiltonianEigenvectors}.
Otherwise, some other monomer frontier orbitals must contribute to the pair wavefunction,
and our simplifying assumption of two interacting orbitals fails.
For all examples results here, the eigenvectors are well represented by this form. 

To input, multiply, and integrate cube files,
we employ a set of Python scripts called {\tt cube\_tools} 
(available in GitHub at https://github.com/funkymunkycool/Cube-Toolz). 
We have modified these scripts to eliminate unnecessary text output,
so they can be used more conveniently as command line tools.
Details about our modified scripts can be found in Supplemental Information. 

The resulting ``Frontier Orbital Numerical Projection'' method 
is computationally efficient, simple, and versatile.
In the next section, we use it to investigate 
how hopping matrix elements and onsite energy shifts vary 
with the identity and placement of interacting molecules. 

\section{Scans}
Scans explore how the matrix hopping element varies 
as we vary the relative location of the two molecules in a pair.
In the scans presented here, one molecule in a pair
is rigidly translated or rotated while the other is held fixed.
Along the scan, the energy splitting $\Delta E$ and eigenvector coefficient $\alpha$ 
are calculated from DFT results and used to calculate $t$.

To prepare the sequence of DFT input files for a scan,
we first optimize the geometry of each monomer.
Next we prepare a {\tt .com} file (Gaussian input file)
in which the two monomer centers of mass
are placed in a given initial geometry.
We then generate a sequence of {\tt .com} files 
by applying scripts that transform the initial {\tt .com} file
by moving one of the two monomers.
The resulting {\tt .com} files are executed by Gaussian
to produce {\tt .log} and {\tt .chk} files, 
which provide the necessary information to compute $t$,
as described further below.

\begin{figure}[htbp]
\begin{center}
\includegraphics[width=0.4\textwidth]{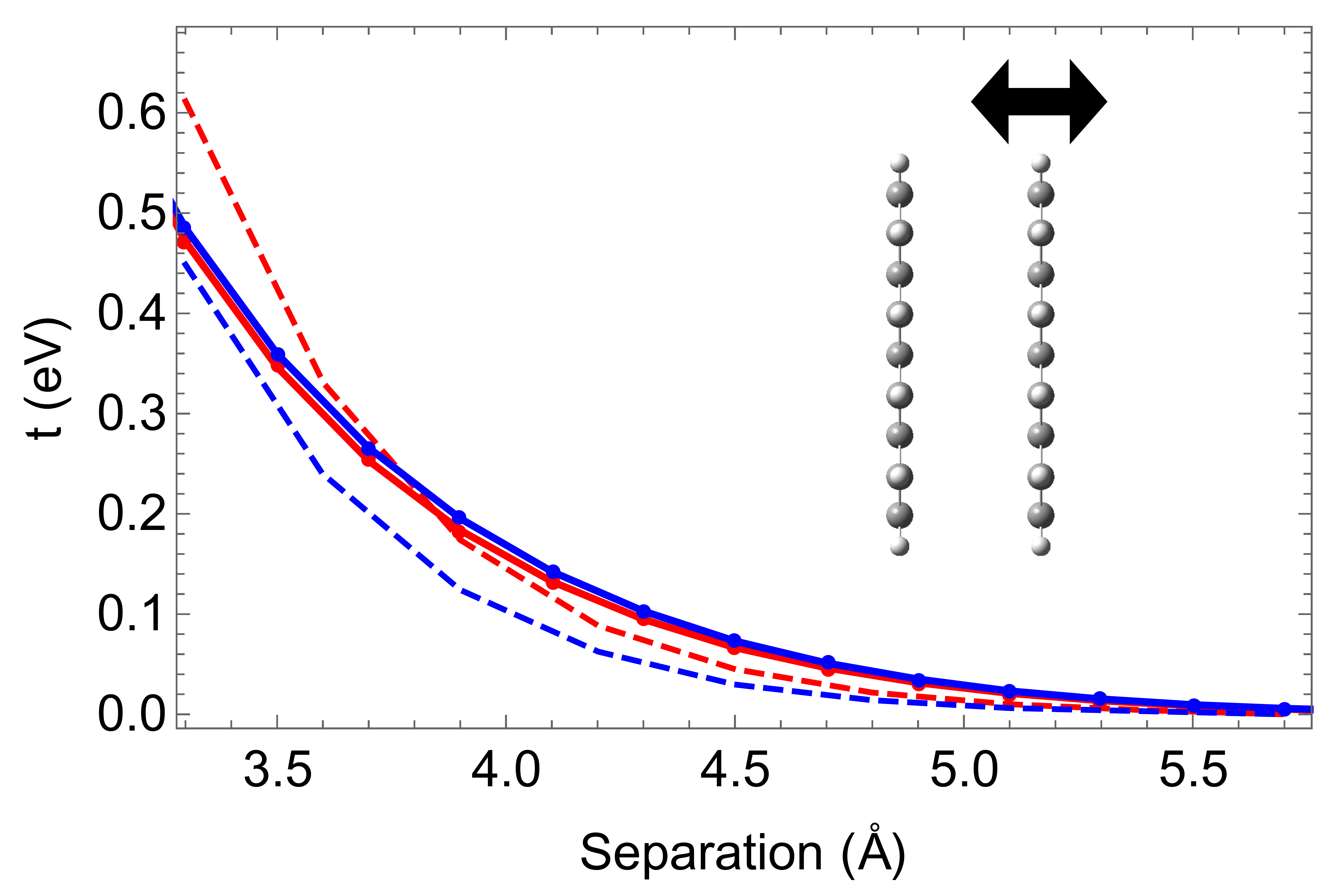}
\caption{$t$ value for hopping between HOMO (red) and LUMO (blue) of each tetracene, 
from our results (solid lines) and Ref.\ \citenum{Coropceanu2007} (dashed lines).}
\label{fig:TetraceneScan}
\end{center}
\end{figure}

To test our new method, we first compare to previous results 
for simple scans of pairs consisting of two identical molecules. 
In the first example, the two molecules are symmetrically placed,
so that the ESD method suffices.
Fig.\ \ref{fig:TetraceneScan} presents results of Coropceanu et al.,
who performed a translation scan for 
two tetracene molecules in a parallel arrangement,
and computed $t$ for hybridization of the monomer HOMOs and LUMOs
\cite{Coropceanu2007}.
 
Evidently, our results and those of Ref.\ \citenum{Coropceanu2007} are quite comparable,
differing in only minor respects:
1) we find $t$ for hopping between tetracene HOMOs and LUMOs are essentially equal,
while Ref.\  \citenum{Coropceanu2007} finds that the HOMO $t$ values are slightly higher;
and 2) we find a slightly larger range for the hopping,
which may reflect a more generous basis set in our DFT calculations. 

\begin{figure}[htbp]
\begin{center}
\includegraphics[width=0.4\textwidth]{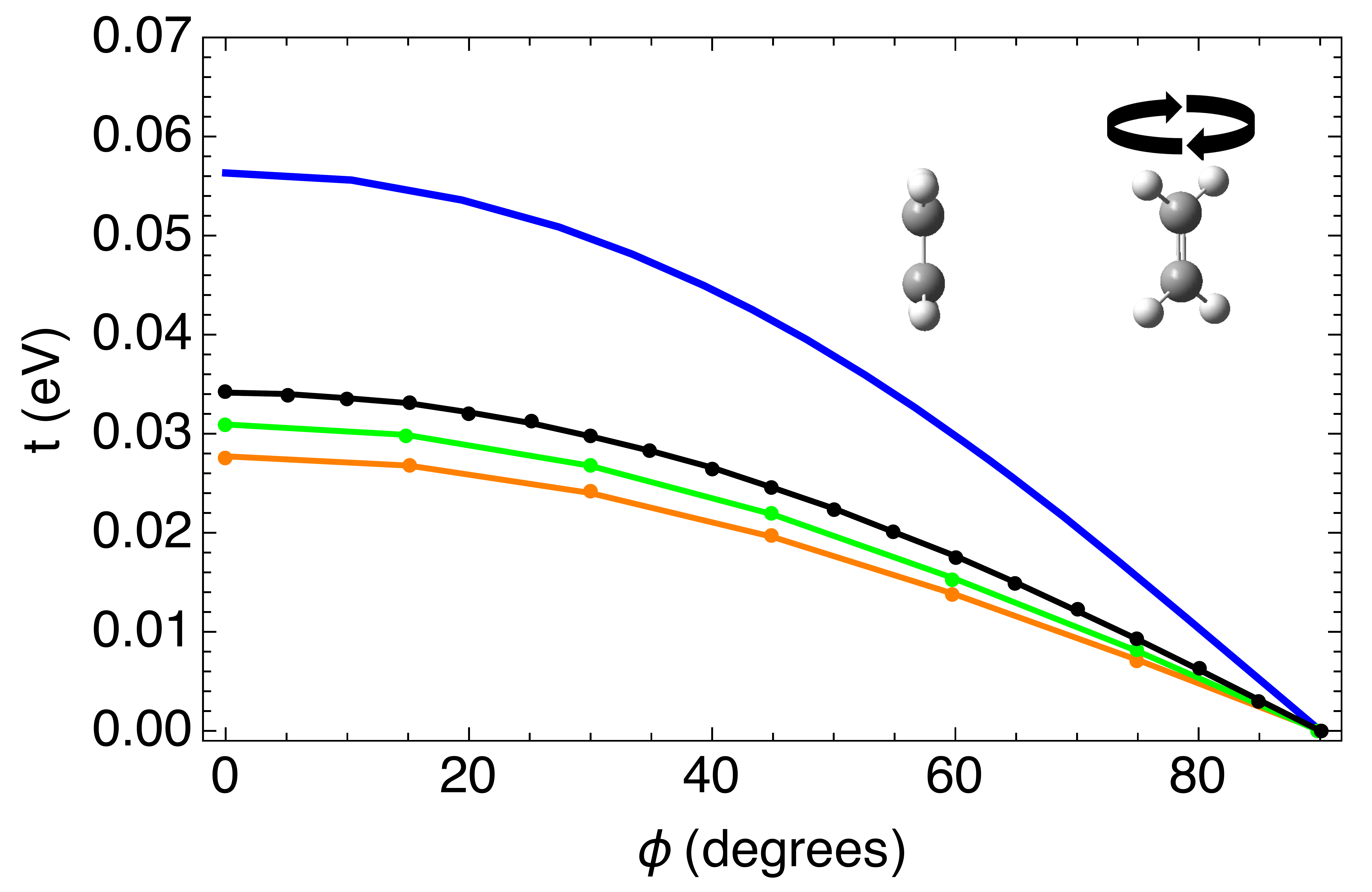}
\caption{HOMO $t$ value for ethylene pair versus rotation angle $\phi$ at separation 5 \AA, 
from Ref.\ \citenum{Valeev2006} (blue), present work (black), 
Ref.\ \citenum{Kirkpatrick2008} MOO method (green) and projective method (orange).}
\label{fig:EthyleneRotation}
\end{center}
\end{figure}

In the second example,
Valeev et al.\ calculated $t$ for hopping between HOMOs 
of two ethylene molecules at a separation of 5 \AA, 
in which one ethylene was rotated around its double bond\cite{Valeev2006}.
This rotation breaks the mirror symmetry,
so that the ESD method cannot be used\cite{Valeev2006}.
(Valeev et al.\ is not explicit about how their $t$ values were computed,
but presumably this was done analytically, 
by computing Fock operator matrix elements
in the basis set used for the SCF calculations.)
Fig.\ \ref{fig:EthyleneRotation} presents their results,
compared to our results for the same scan.

Kirkpatrick also calculated $t$ for the very same scan\cite{Kirkpatrick2008},
using two different approaches.
In the first approach, Ref.\ \citenum{Kirkpatrick2008} 
analytically computed matrix elements of the Fock operator and hence $t$.
The second approach, which is limited by the ZINDO approximation,
quantum calculations are required only of individual molecules and not of a pair,
constructs hybrid bonding and anti-bonding orbitals 
from the orbitals of the isolated molecules.
Results from both approaches appear in Fig.\ \ref{fig:EthyleneRotation}.

Our Frontier Orbital Numerical Projection method
can likewise calculate $t$ for this ethylene pair rotation scan. 
To proceed, we use the Gaussian utility {\tt cubegen}
to generate cube files for the frontier orbitals of the pair
as well as the constituent frontier orbitals of the two monomers,
for each pair geometry along the scan.
The cube files are generated with the same origin, axes, and mesh size,
to facilitate their use in computing the necessary inner products
to project the pair orbital onto the constituent monomer orbitals.
Information about our scripts that automate these overlap calculations
can be found in Supplemental Information. 

As an example, for the ethylene pair rotation scan at a rotation angle of 10 degrees, 
calculating as described above the overlaps of each monomer HOMO
with the pair HOMO and HOMO-1, 
generates eigenvector expansion coefficients 
\begin{equation}
\left\lbrack
\begin{array}{ccc}
  e^+_1 & e^-_1\\
  e^+_2 & e^-_2 \\
\end{array}
\right\rbrack = \left\lbrack
\begin{array}{ccc}
  0.721 & 0.693 \\
  -0.695 & 0.719 \\
\end{array}
\right\rbrack
\label{eqn:EthyleneMatrix}
\end{equation}

The coefficient matrix Eqn.\ \ref{eqn:EthyleneMatrix} is almost perfectly antisymmetric,
as expected from the analytical form Eqn.\ \ref{eqn:HamiltonianEigenvectors}
for an interacting $2 \times 2$ system.
Inevitably, the coefficient values do not quite satisfy the expected relations
$e^+_1 = e^-_2$ and $e^+_2 = -e^-_1$,
so that $\alpha$ obtained from $\alpha = e^-_1/e^-_2$
does not quite equal $\alpha = -e^+_2/e^+_1$.
Consequently, we average the two results, to obtain
 \begin{equation}
	\alpha=(1/2)\left(\left|e^-_1/e^-_2\right|
	+\left|e^+_2/e^+_1\right|\right) 
	\label{eqn:AlphaFromMatrix}
\end{equation}

The hopping matrix element $t$ versus rotation angle $\phi$ can then be calculated 
from $\alpha(\phi)$, the pair HOMO and HOMO-1 energies, and Eqn.\ \ref{eqn:Finalt}.
Fig.\ \ref{fig:EthyleneRotation} shows that results from our method (black)
are in good agreement with both methods of Ref.\ \citenum{Kirkpatrick2008} (green and orange),
while results from Ref.\ \citenum{Valeev2006} (blue) are systematically about two times larger. 

Having tested our Frontier Orbital Numerical Projection method in this way,
we now use it to investigate hopping matrix elements 
for moieties that appear in organic semiconductors
under development for photovoltaic materials.
In what follows, we select pairs of molecules to highlight the importance 
of the energy, symmetry, and relative placement of the interacting orbitals.

\section{Thiophene translation scan} 

\begin{figure*}
\begin{center}
\includegraphics[width=0.9 \textwidth]{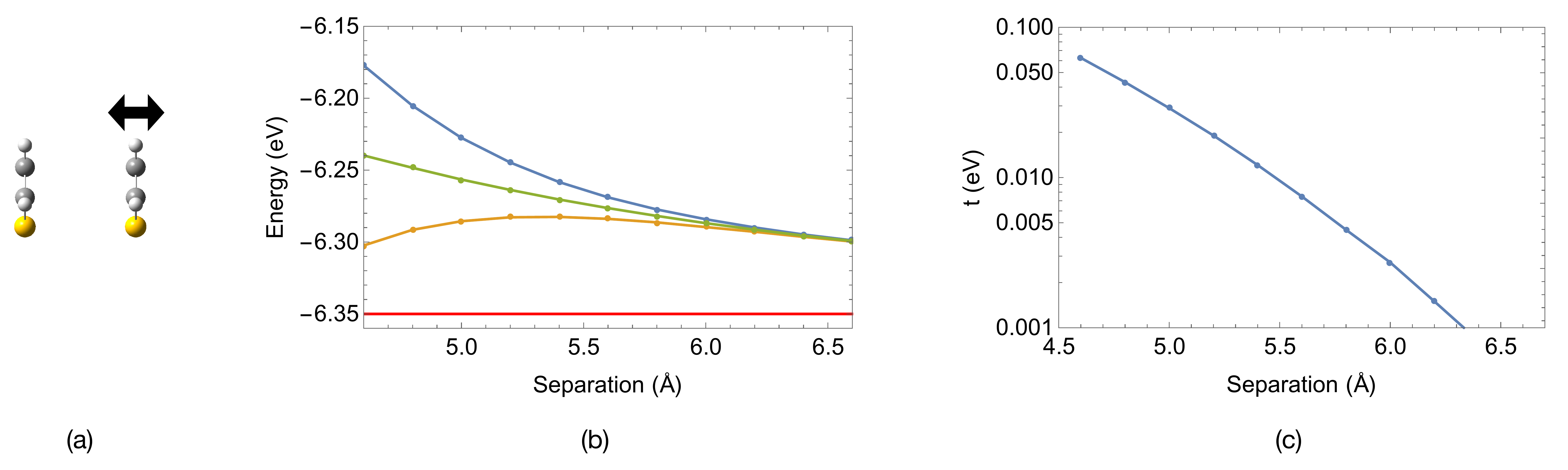}
\caption{(a): Schematic of thiophene translation scan. 
(b): Pair HOMO (blue) and HOMO-1 (orange) energies and average of the two (green) versus separation. 
Red asymptote is HOMO of isolated monomer. (c): Resulting $t$ value versus separation.} 
\label{fig:ThThTransScan}
\end{center}
\end{figure*}

For our first example scans, we investigate a pair of thiophene molecules,
thiophene being a well-studied constituent of many organic semiconductors.
Intuitively, we expect large $t$ values for thiophenes placed face to face, 
as in Fig.\ \ref{fig:ThThTransScan}.

We first carry out a scan with respect to the separation between the thiophenes.
Throughout this scan, the monomer HOMOs hybridize to form the pair HOMO and HOMO-1,
while the monomer LUMOs hybridize to form the pair LUMO and LUMO+1.
For this scan, the two molecules remain symmetrically placed,
so $\alpha$ is equal to unity throughout the scan,
and $t$ can be computed from the splitting alone as $t = \Delta E/2$. 

Fig.\ \ref{fig:ThThTransScan} shows the scan geometry,
the pair HOMO and HOMO-1 energies versus distance,
and the resulting values of $t$.
Qualitatively similar results with slightly higher $t$ values are found for hopping between the LUMOs. 

As Fig.\ \ref{fig:ThThTransScan}(c) shows,
$t$ decreases approximately exponentially with distance,
reflecting the approximately exponential tails of the interacting monomer orbitals.
This behavior is consistent with distance scans for other moieties
\cite{Kwon:tt,Valeev2006}.

 Fig.\ \ref{fig:ThThTransScan}(b) shows that the pair HOMO and HOMO-1 energies 
 are essentially degenerate beyond 6.5 \AA\ separation. 
However, at this separation the HOMO and HOMO-1 remain significantly higher 
than the energy of the isolated monomer HOMO
(red line in Fig.\ \ref{fig:ThThTransScan}(b)).
The HOMO and HOMO-1 pair energies
continue decreasing with distance until roughly 30 \AA.
This slow approach to truly isolated molecules does not arise from hybridization,
but long-range electrostatic interactions between the molecules,
each of which perturbs the HOMO of the other. 

At long distances, the dominant interactions between neutral molecules 
should be dipole-dipole interactions, as remarked by Valeev et al.\cite{Valeev2006}.
We can investigate this by analyzing the onsite energy shift 
of the pair orbitals with respect to the isolated limit.

To calculate the onsite energy shift,
we average the pair HOMO and HOMO-1 energies,
and subtract the HOMO energy of an isolated thiophene.
Because the pair consists of identical molecules symmetrically placed,
the average of the pair HOMO and HOMO-1
must equal the onsite energy of each of the two monomers,
which are by symmetry equal.

Calculating this average against distance
gives a relationship fit by
Equation \ref{eqn:GreenGraphFittedEquation}.
Beyond a separation of 10 \AA,
the correction is less than 25 percent of the leading $1/R^3$ dependence,
consistent with the interaction energy between two dipoles at a distance $R$.

\begin{equation}
	U(R) = -\frac{51.1004}{R^4}+\frac{21.7046}{R^3}
	\label{eqn:GreenGraphFittedEquation}
\end{equation}

\section{Thiophene rotation scan}

On physical grounds, we expect the value of $t$ to depend 
on the relative orientation of the two interacting monomers, 
which governs how the lobes of the frontier orbitals overlap. 
We explore the dependence of $t$ on orientation
by rotating one of a pair of thiophenes,
at a constant separation distance of 4.2 \AA.

\begin{figure*}
\begin{center}
\includegraphics[width=0.9\textwidth]{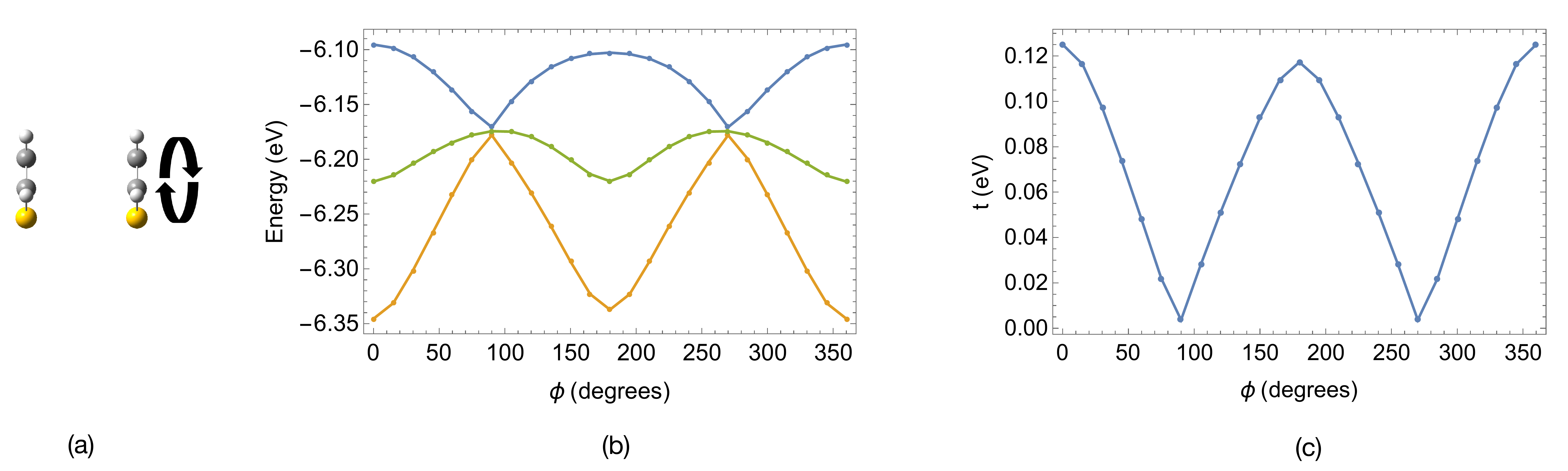}
\caption{(a): Schematic of thiophene rotation scan at 4.2 \AA\ separation. 
(b): Pair HOMO (blue) and HOMO-1 (orange) energies and their average (green) versus $\phi$. 
(c): Resulting t value versus $\phi$.}
\label{fig:ThThRotScanHOMO}
\end{center}
\end{figure*}

Throughout the scan, the monomer HOMOs hybridize to form the pair HOMO and HOMO-1, 
while the monomer LUMOs hybridize to form the LUMO and LUMO+1.
Fig.\ \ref{fig:ThThRotScanHOMO} shows the energy of the HOMO and HOMO-1 pair orbitals 
and their corresponding value of $t$ throughout the rotation. 
Because the two thiophenes are related throughout the scan
by a two-fold rotational symmetry,
the ESD method again suffices to determine the value of $t$ versus $\phi$.

\begin{figure}[htbp]
\begin{center}
\includegraphics[width=0.4\textwidth]{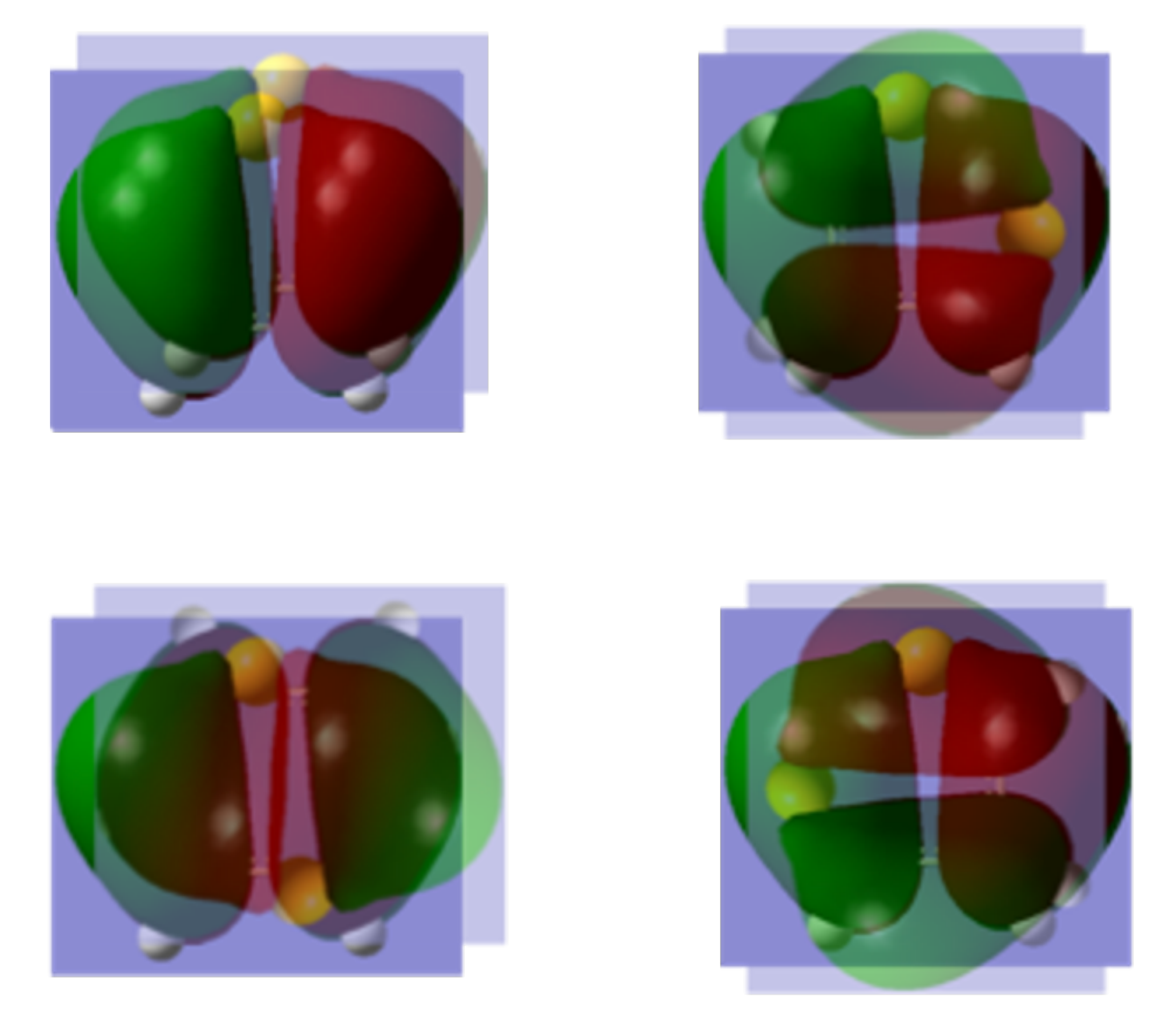}
\caption{Images of overlaying thiophene HOMOs at 0, 90, 180, and 270 degrees.}
\label{fig:ThThHomoOverlaps}
\end{center}
\end{figure}

Fig.\ \ref{fig:ThThHomoOverlaps} displays images 
of how the HOMOs of the two monomers in the pair overlap 
at various rotations throughout the scan.
Evidently, at rotations of zero and 180 degrees,
the proximate lobes on the monomer HOMOs line up consistently,
so that symmetric and antisymmetric linear combinations of the two HOMOs
lead to bonding and antibonding pair orbitals.
Whereas, at 90 and 270 degrees,
the proximate lobes are orthogonally oriented,
so that regardless of the choice of sign 
for the amplitudes of the two monomer HOMOs,
no consistent pattern of nodes or ``joins''
between the two HOMOs can be established.

This state of affairs is reflected 
in the periodic dependence of $t$ on $\phi$.
At 0, 180, and 360 degrees
the interacting lobes directly face each other,
maximizing hybridization and thus $t$. 
At 90 and 270 degrees, hybridization hardly lowers the energy of the pair HOMO-1.

\begin{figure}
     \centering
     \begin{subfigure}[b]{0.4\textwidth}
         \centering
         \includegraphics[width=\textwidth]{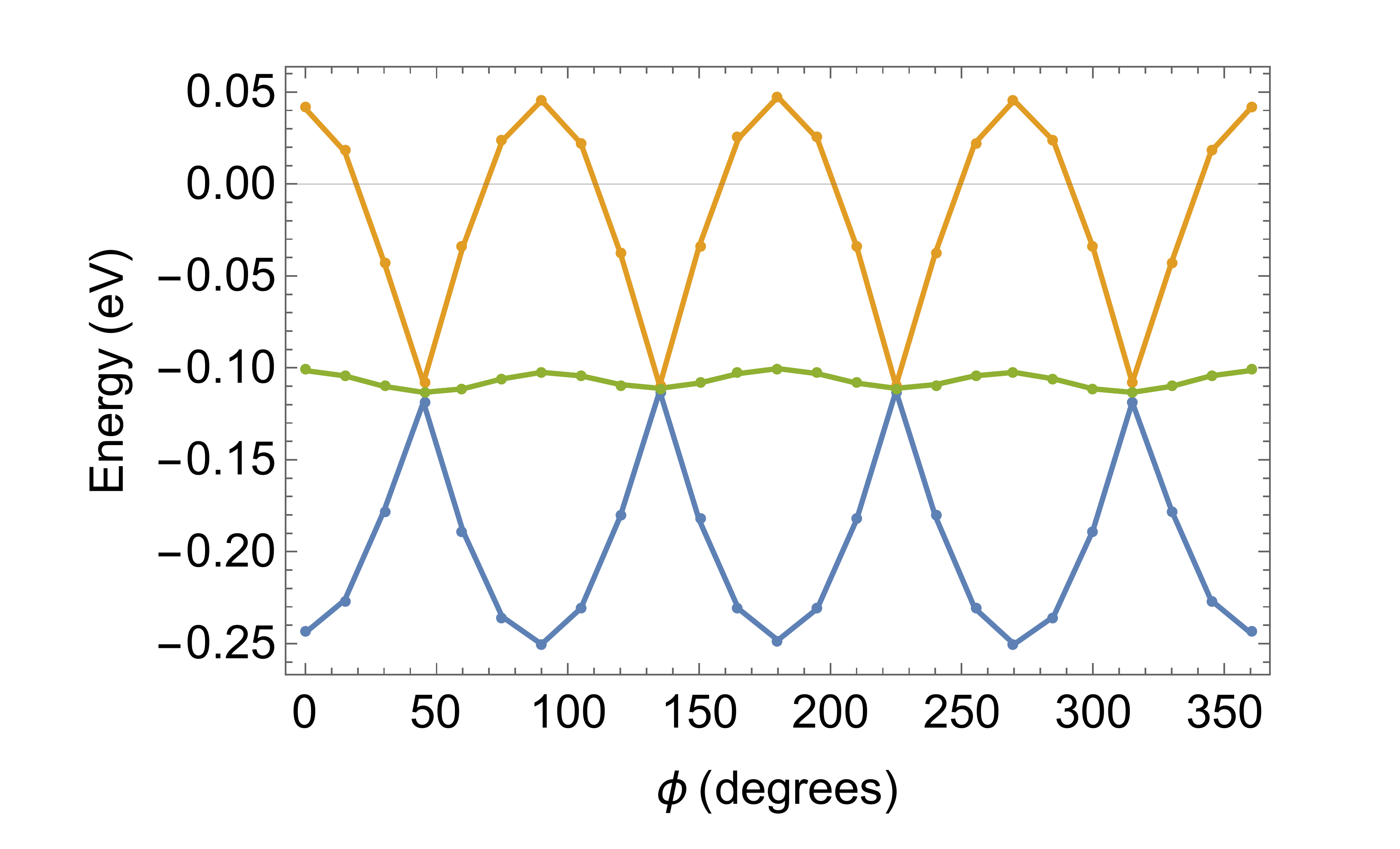}
         \caption{}
         \label{fig:ThRotScanLUMOEnerg}
     \end{subfigure}
     \hfill
     \begin{subfigure}[b]{0.4\textwidth}
         \centering
         \includegraphics[width=\textwidth]{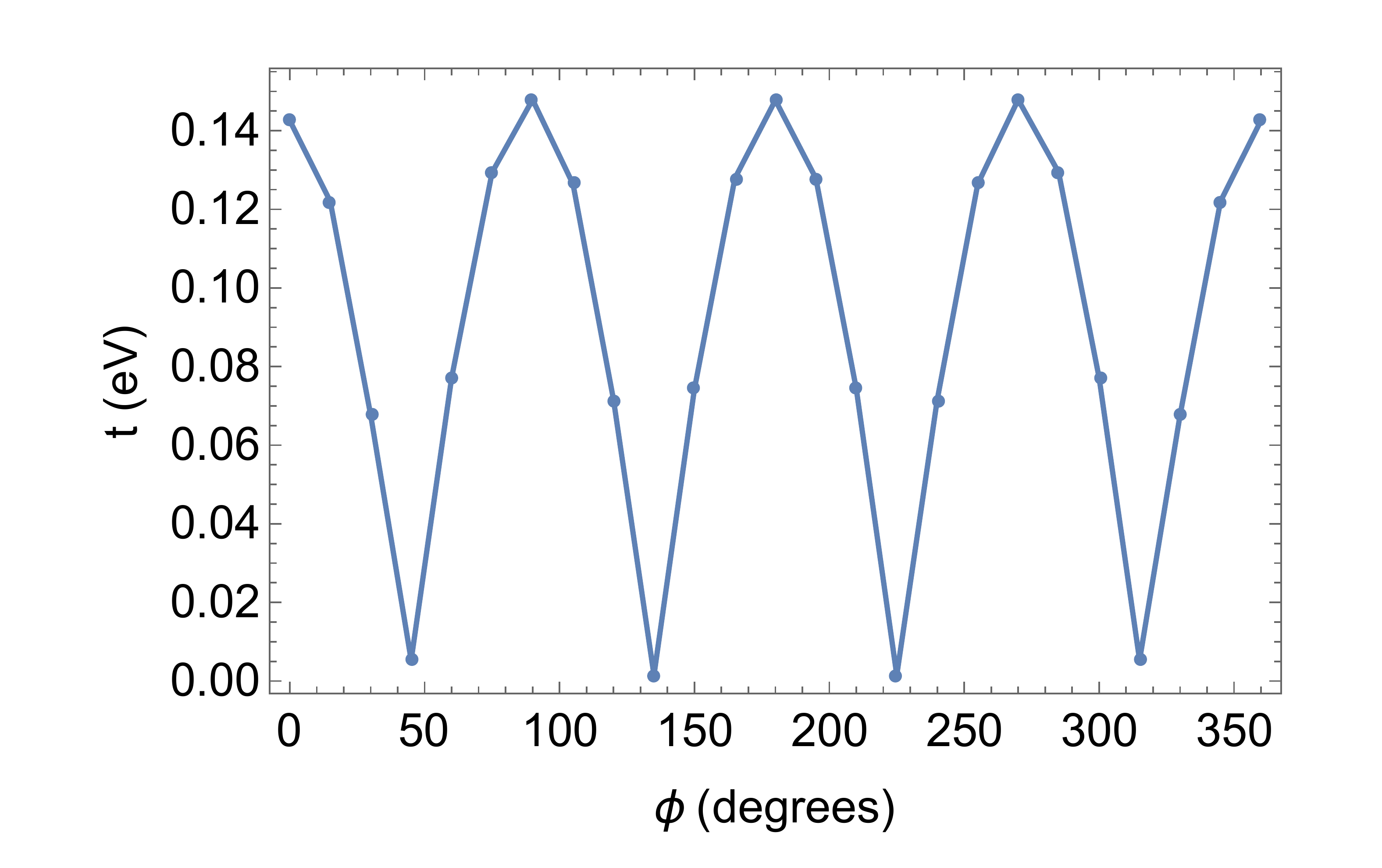}
         \caption{}
         \label{fig:ThRotScanLUMOt}
     \end{subfigure}
       \caption{a): Pair LUMO (blue) and LUMO+1 (orange) energies 
			for thiophene and thiophene scan at 4.2 \AA\ separation 
			and their average (green) versus $\phi$. (b): Resulting t value versus $\phi$.}
        \label{fig:ThThRotScanLUMO}
\end{figure}

The thiophene LUMO has a different pattern of lobes than the HOMO,
resulting in a different variation of the energy shifts and $t$ values versus $\phi$, 
as shown in Fig.\ \ref{fig:ThThRotScanLUMO}.

\begin{figure}[htbp]
\begin{center}
\includegraphics[width=0.4\textwidth]{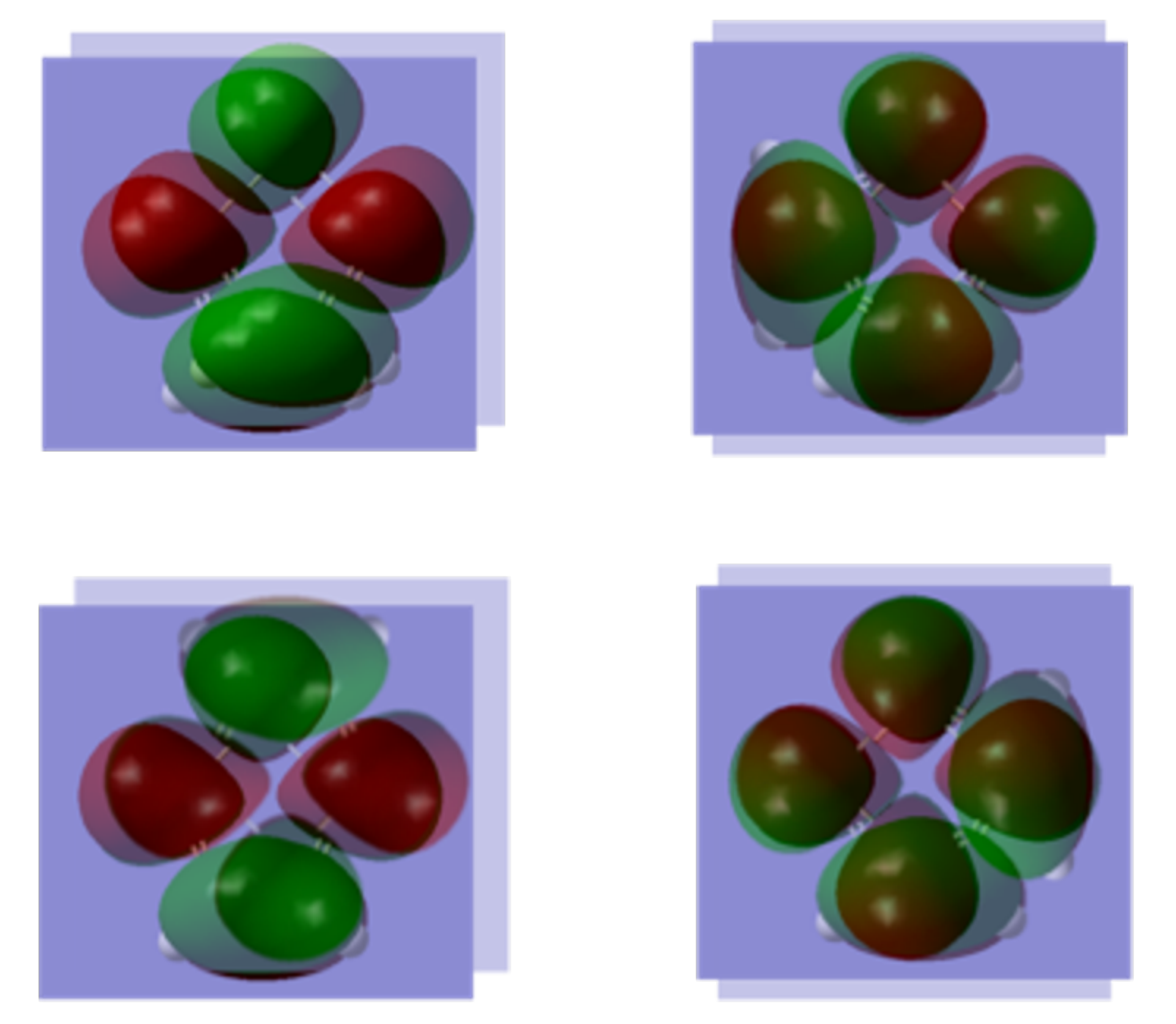}
\caption{Images of overlaying thiophene LUMOs at 0, 90, 180, 270 degrees.}
\label{fig:ThThLumoOverlaps}
\end{center}
\end{figure}

The periodicity of energy and $t$ versus rotation is determined 
by the arrangement of lobes on the interacting frontier orbitals.
The schematics in Fig.\ \ref{fig:ThThLumoOverlaps} illustrate 
the four opportunities for the frontier orbitals to line up for maximum hybridization,
which results in four maxima in $t(\phi)$.

\section{Benzene pairs}

\begin{figure}[htbp]
\begin{center}
\includegraphics[width=0.4\textwidth]{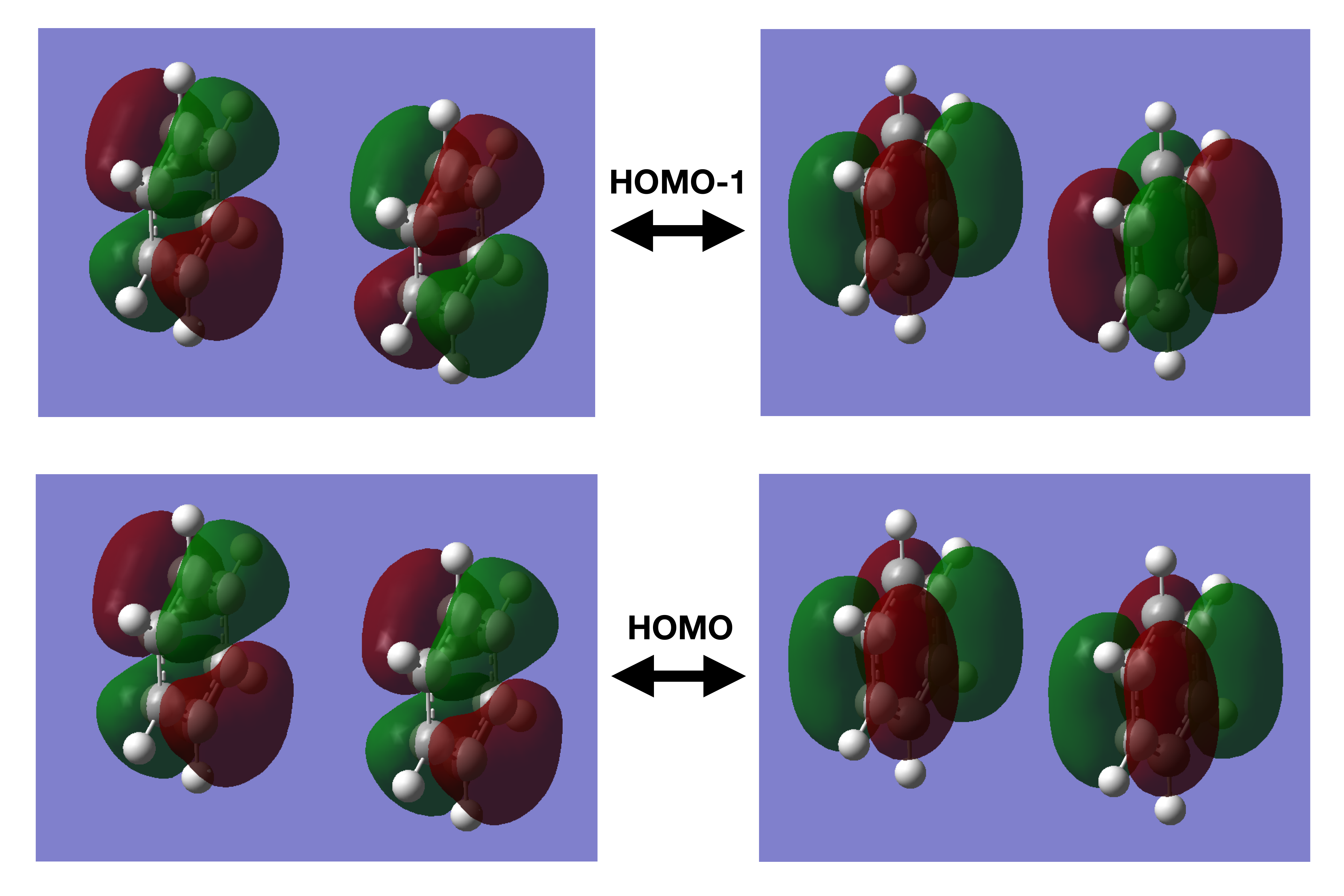}
\caption{Benzene pair orbitals with the HOMO-1 and HOMO energy at 4.2 \AA\ separation.}
\label{fig:BenzeneOrbitals}
\end{center}
\end{figure}

Because of its six-fold rotational symmetry,
both the HOMO and LUMO of benzene are doubly degenerate,
which has interesting consequences for hybridization 
between two benzene moieties in close proximity.
For example, symmetrically placed benzene pairs 
also have doubly degenerate frontier orbitals. 
Fig.\ \ref{fig:BenzeneOrbitals} shows the frontier orbitals 
for a pair of face-on benzene molecules;
two different bonding orbitals have the same HOMO-1 energy,
and two different antibonding orbitals have the same HOMO energy. 

\begin{figure}
     \centering
     \begin{subfigure}[b]{0.425\textwidth}
         \centering
         \includegraphics[width=\textwidth]{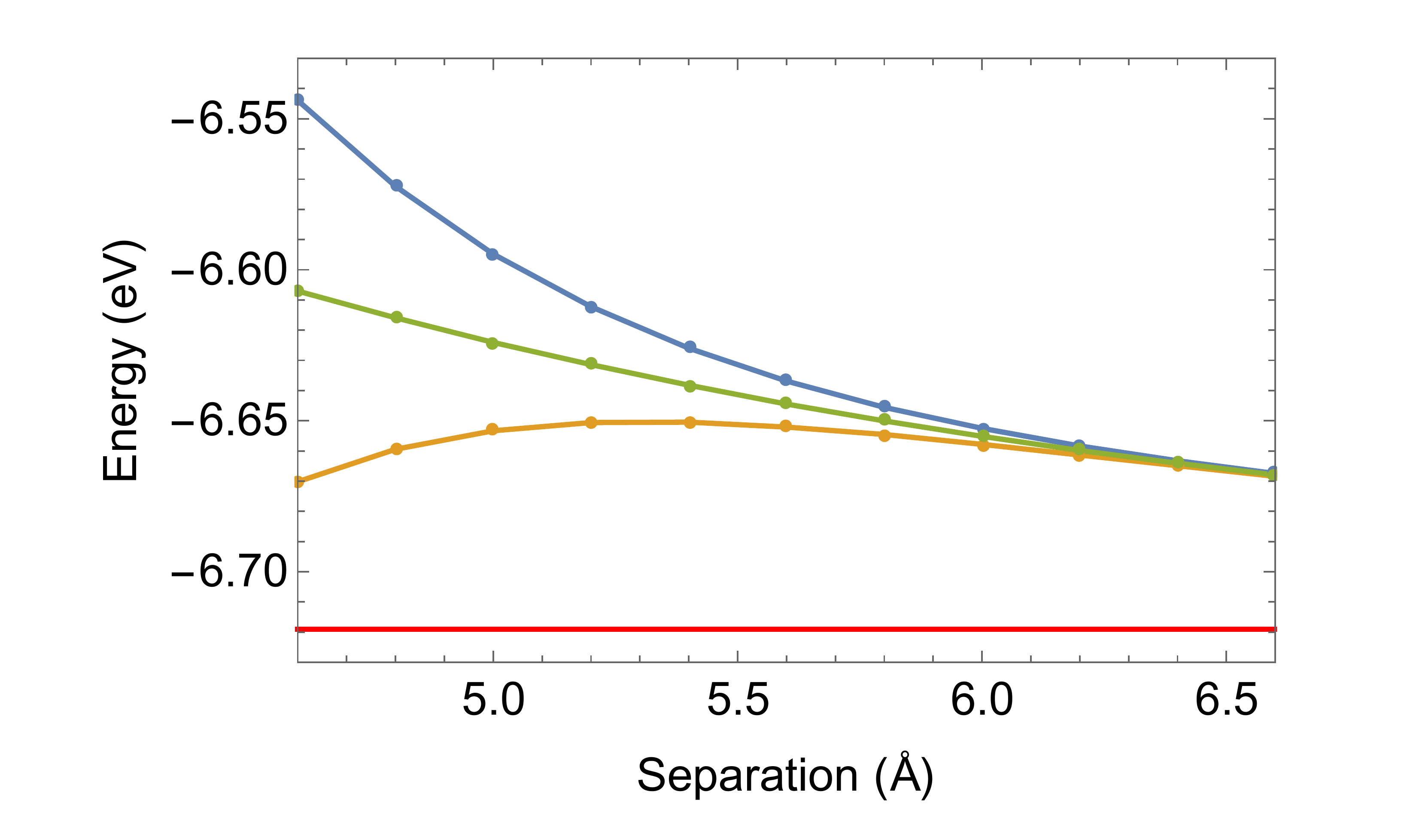}
         \caption{}
         \label{fig:BenzEnergScanTrans}
     \end{subfigure}
     \hfill
     \begin{subfigure}[b]{0.4\textwidth}
         \centering
         \includegraphics[width=\textwidth]{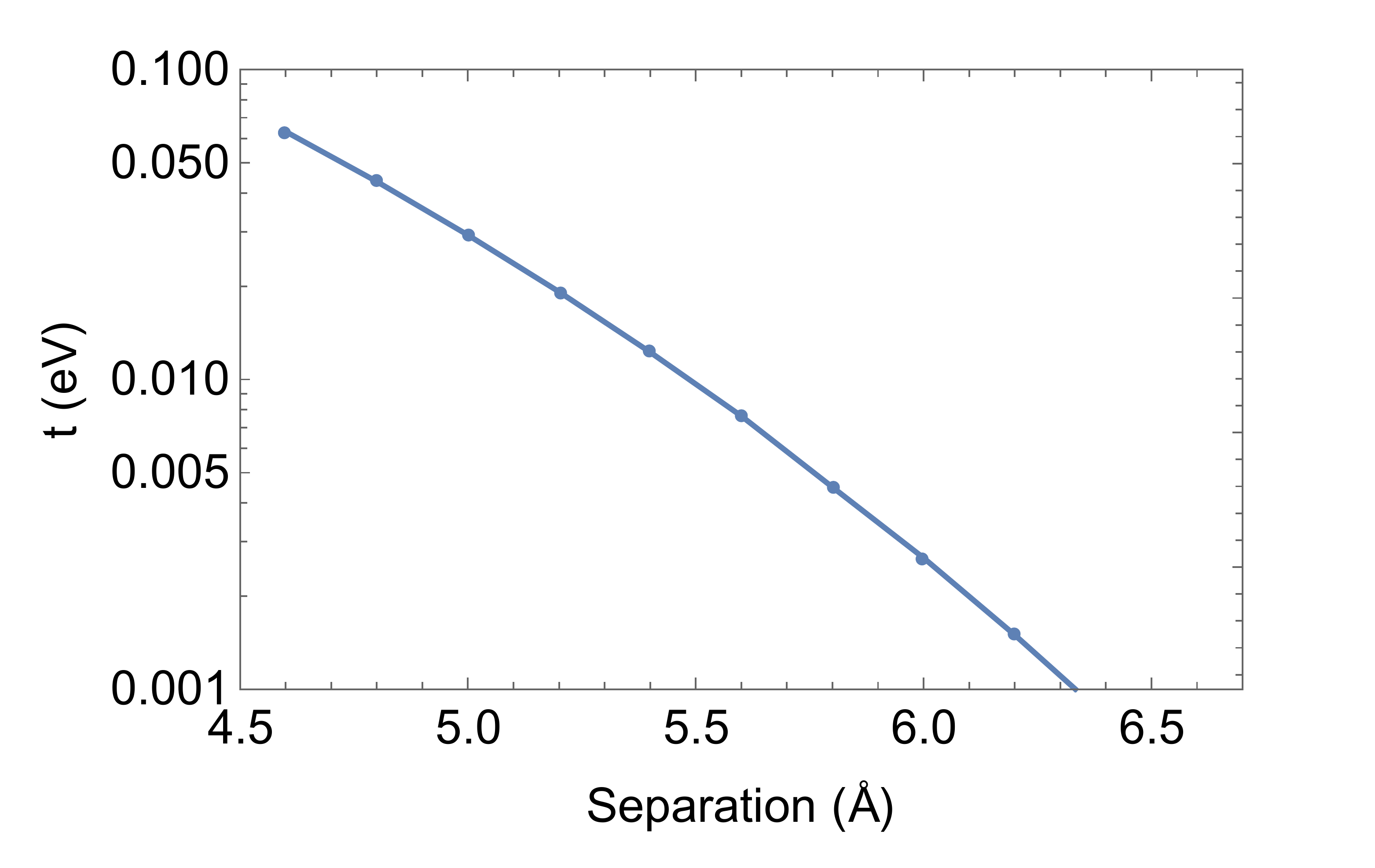}
         \caption{}
         \label{fig:BenztScanTrans}
     \end{subfigure}
       \caption{(a): Pair HOMO (blue) and HOMO-1 (orange) energies for benzene and benzene scan and their average (green) versus separation. Red asymptote is HOMO of isolated monomer. (b): Resulting $t$ value versus separation.}
        \label{fig:BenzTransScan}
\end{figure}

Fig.\ \ref{fig:BenzTransScan} shows results for a translation scan of two benzenes,
performed as for thiophene in Section V. 
Because the molecules are symmetrically placed,
the ESD method suffices to compute $t$.
Throughout the scan, the pair HOMO and HOMO-1 are doubly degenerate.

The graph of $t$ versus separation in Fig.\ \ref{fig:BenzTransScan} 
is strikingly similar to the results for thiophene.
As for thiophene, the benzene pair HOMO and HOMO-1 
approach one another much faster 
than they approach the isolated benzene monomer HOMO,
suggesting the presence of long-range interactions
that shift the energy of pair orbitals.

The sixfold symmetry of benzene and resulting degenerate frontier orbitals
has a surprising consequence for $t(\phi)$.
The two degenerate HOMOs on one benzene can be combined to produce an orbital 
that aligns optimally with the other benzene throughout the entire scan,
minimizing the energy of the bonding orbital. 
\begin{figure}[htbp]
\begin{center}
\includegraphics[width=0.4\textwidth]{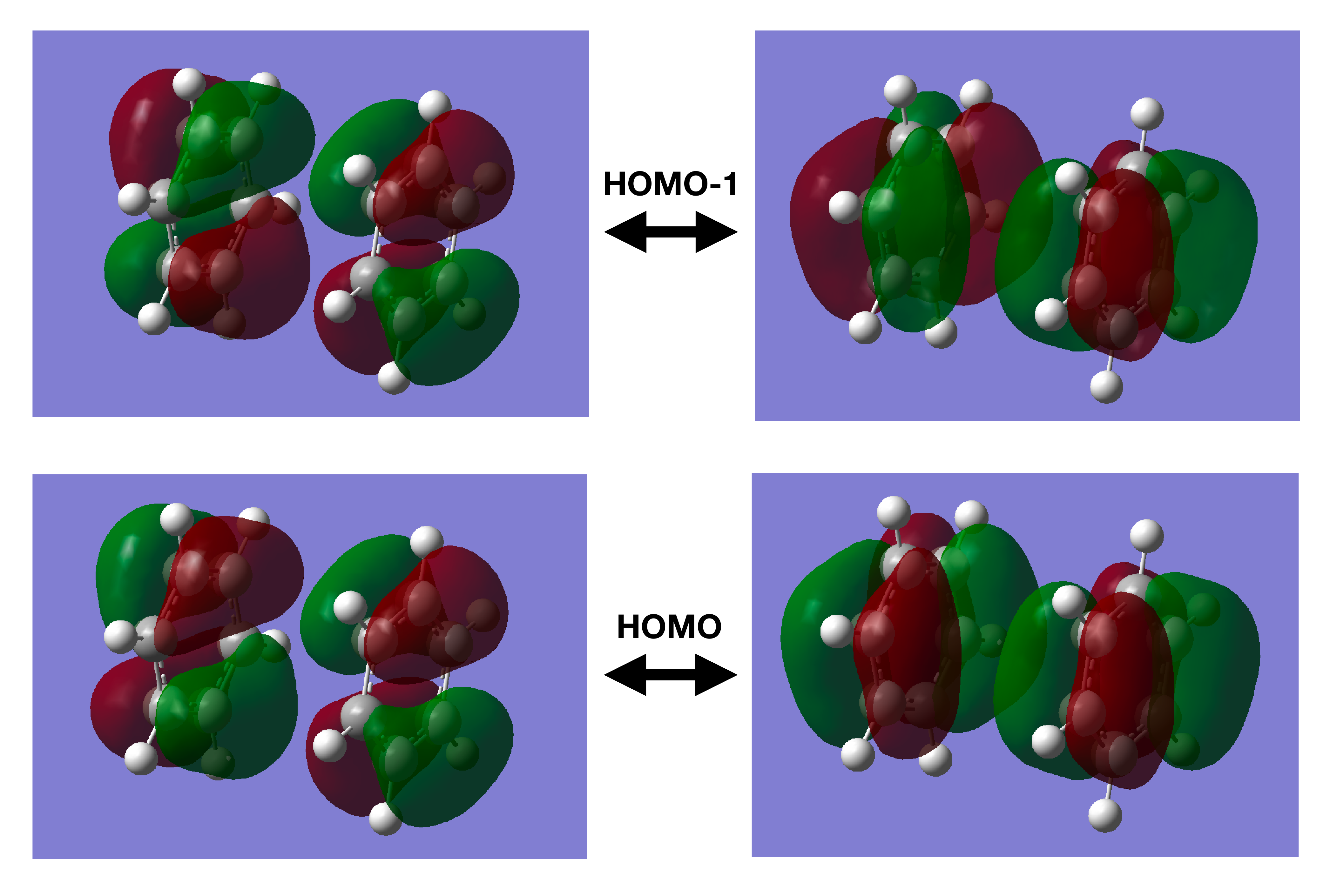}
\caption{Benzene pair orbitals with the HOMO-1 and HOMO energy 
at 4.2 \AA\  separation and 30 degrees rotation.}
\label{fig:BenzOrbitalsRot}
\end{center}
\end{figure}

Fig.\ \ref{fig:BenzOrbitalsRot} shows bonding and antibonding orbitals
for a benzene pair rotated by 30 degrees
--- which are identical to the orbitals of the unrotated pair depicted in Fig.\ \ref{fig:BenzeneOrbitals}.
As a result, for a benzene pair 
the HOMO-1, HOMO, LUMO, and LUMO+1 energies,
as well as the corresponding $t$ values,
are completely independent of rotation.

\section{Thiophene-pyrrole pairs}

The ESD method does not work when the two interacting monomers are different.
Our new approach works for such cases,
which we illustrate here for thiophene and pyrrole. 
The HOMO and LUMO of thiophene and pyrrole 
have similar shapes but rather different energies,
which leads to asymmetric bonding and antibonding pair wavefunctions. 

Although Fig.\ \ref{fig:ThPyrOrbitals} shows the HOMOs
of thiophene and pyrrole forming the HOMO and HOMO-2 of the pair,
this is only true at separations closer than 4.2 \AA,
when the hybridization between the monomer HOMOs is strong enough to
lower the bonding orbital energy to the HOMO-2 position. 
Throughout the scan at larger separations (beyond 4.2 \AA),
the monomer HOMOs hybridize to form the pair HOMO and HOMO-1.

\begin{figure}
     \centering
     \begin{subfigure}[b]{0.4\textwidth}
         \centering
         \includegraphics[width=\textwidth]{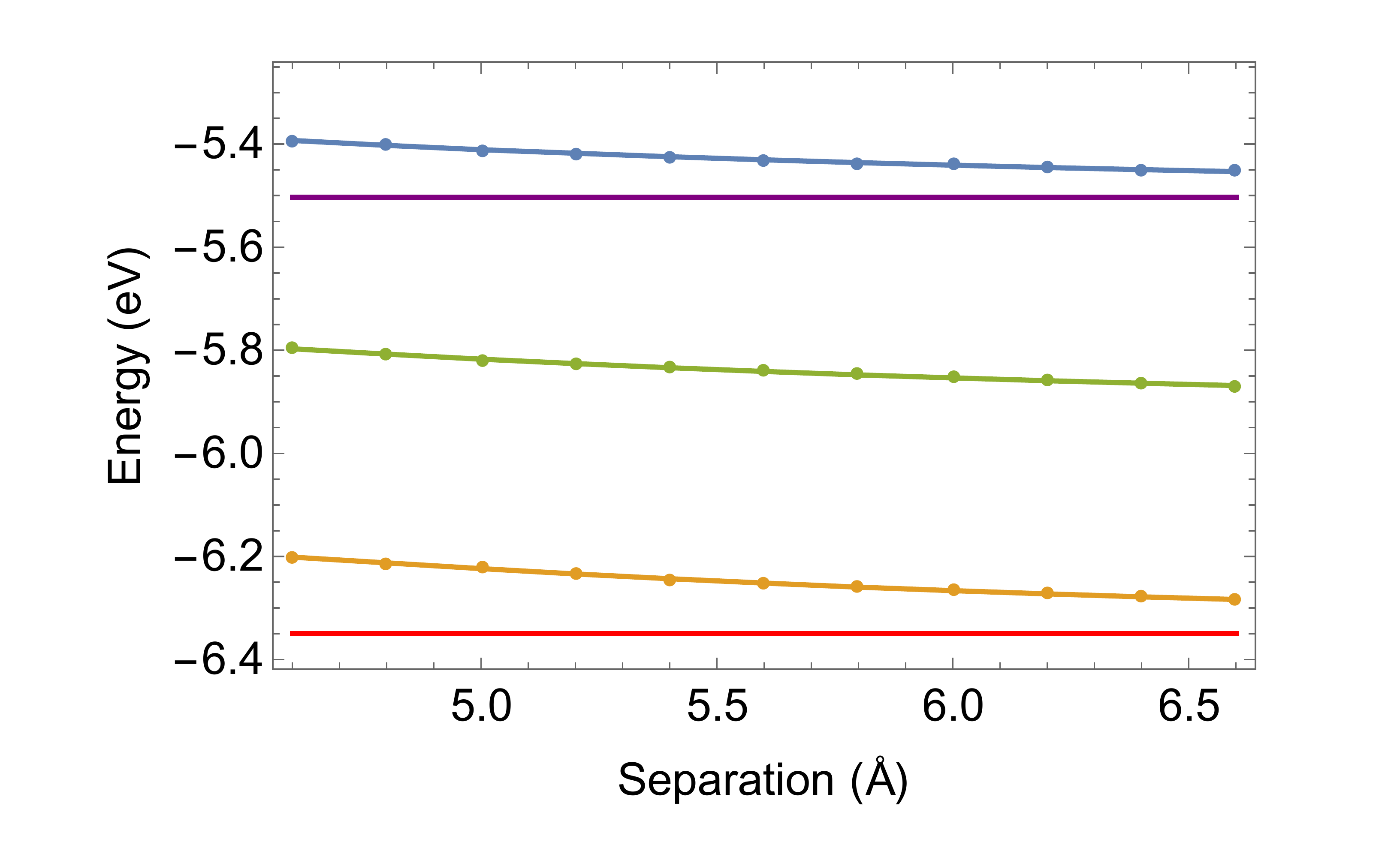}
         \caption{}
         \label{fig:ThioPyrEnergTrans}
     \end{subfigure}
     \hfill
     \begin{subfigure}[b]{0.4\textwidth}
         \centering
         \includegraphics[width=\textwidth]{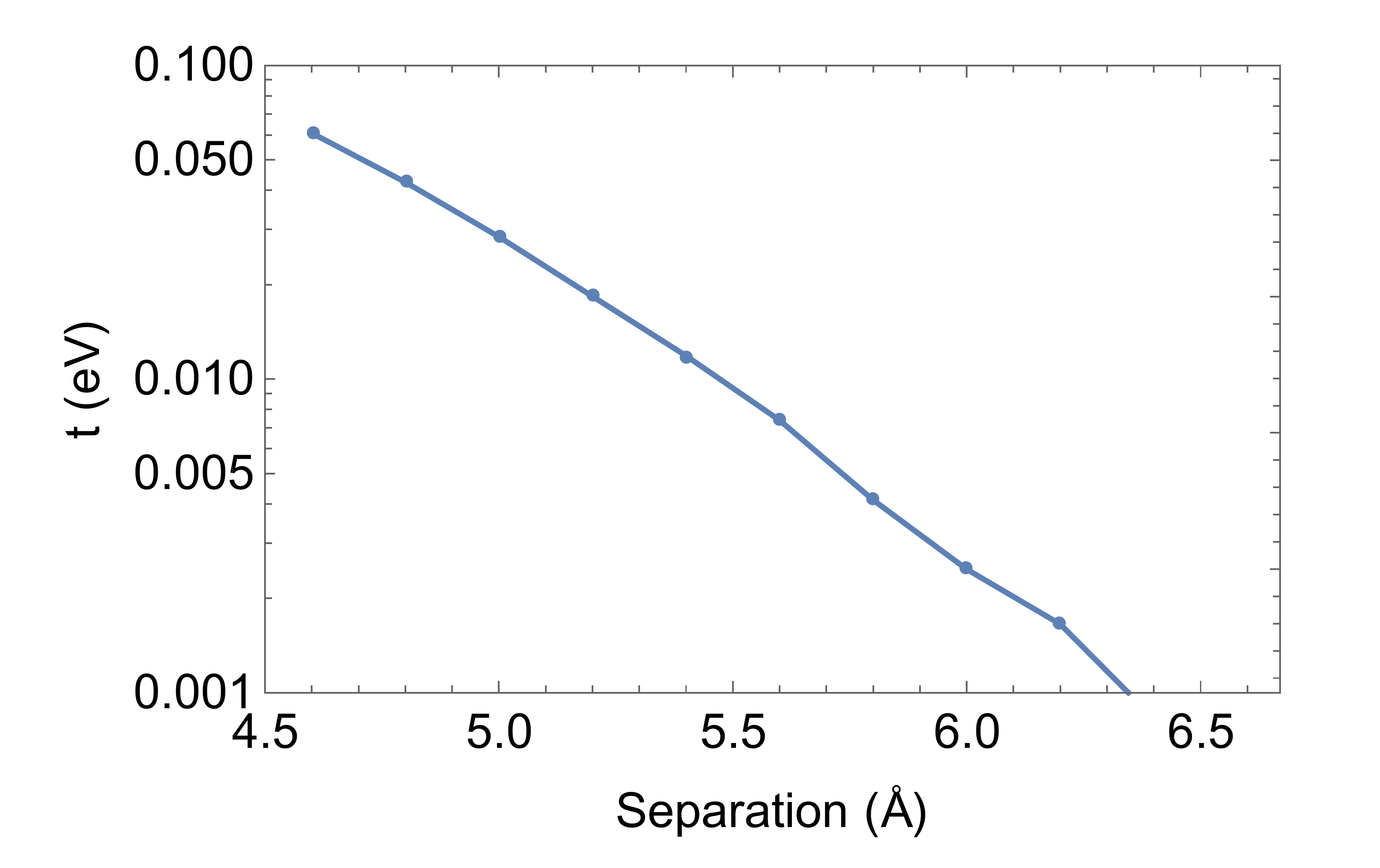}
         \caption{}
         \label{fig:ThioPyrtTrans}
     \end{subfigure}
       \caption{(a): Pair HOMO (blue) and HOMO-1 (orange) energies for thiophene and pyrrole scan 
and their average (green) versus separation. 
Red asymptote is HOMO of isolated thiophene and purple asymptote is HOMO of isolated pyrrole. 
(b): Resulting $t$ value versus separation.}
        \label{fig:ThioPyrTrans}
\end{figure}

Fig.\ \ref{fig:ThioPyrTrans}(a) shows the energies of the pair orbitals 
formed by the thiophene and pyrrole HOMOs throughout the scan.
Fig.\ \ref{fig:ThioPyrTrans}(b) displays the results for $t$ versus separation
obtained using the Frontier Orbital Numerical Projection method.

In Fig.\ \ref{fig:ThioPyrTrans}(a), the pair orbital energies
approach the values for the isolated monomers as the separation distance increases.
The resulting $t$ versus distance is rather similar
to results for the thiophene-thiophene scan,
with similar magnitudes and characteristic exponential decay length.  

Rotation scans of a thiophene-pyrrole pair exhibit an interesting feature,
which is hybridization of orbitals other than the HOMO of both moieties,
as a consequence of close energy matches between disparate orbitals.
The pyrrole HOMO-1 energy of -6.34 eV
is nearly degenerate with the thiophene HOMO energy of -6.35 eV.
As a result, these orbitals strongly hybridize 
when the two monomers are correctly oriented, 
so that the lobes on one monomer orbital
are adjacent to lobes on the other orbital
with a consistent sign difference between them.

\begin{figure}[htbp]
\begin{center}
\includegraphics[width=0.4\textwidth]{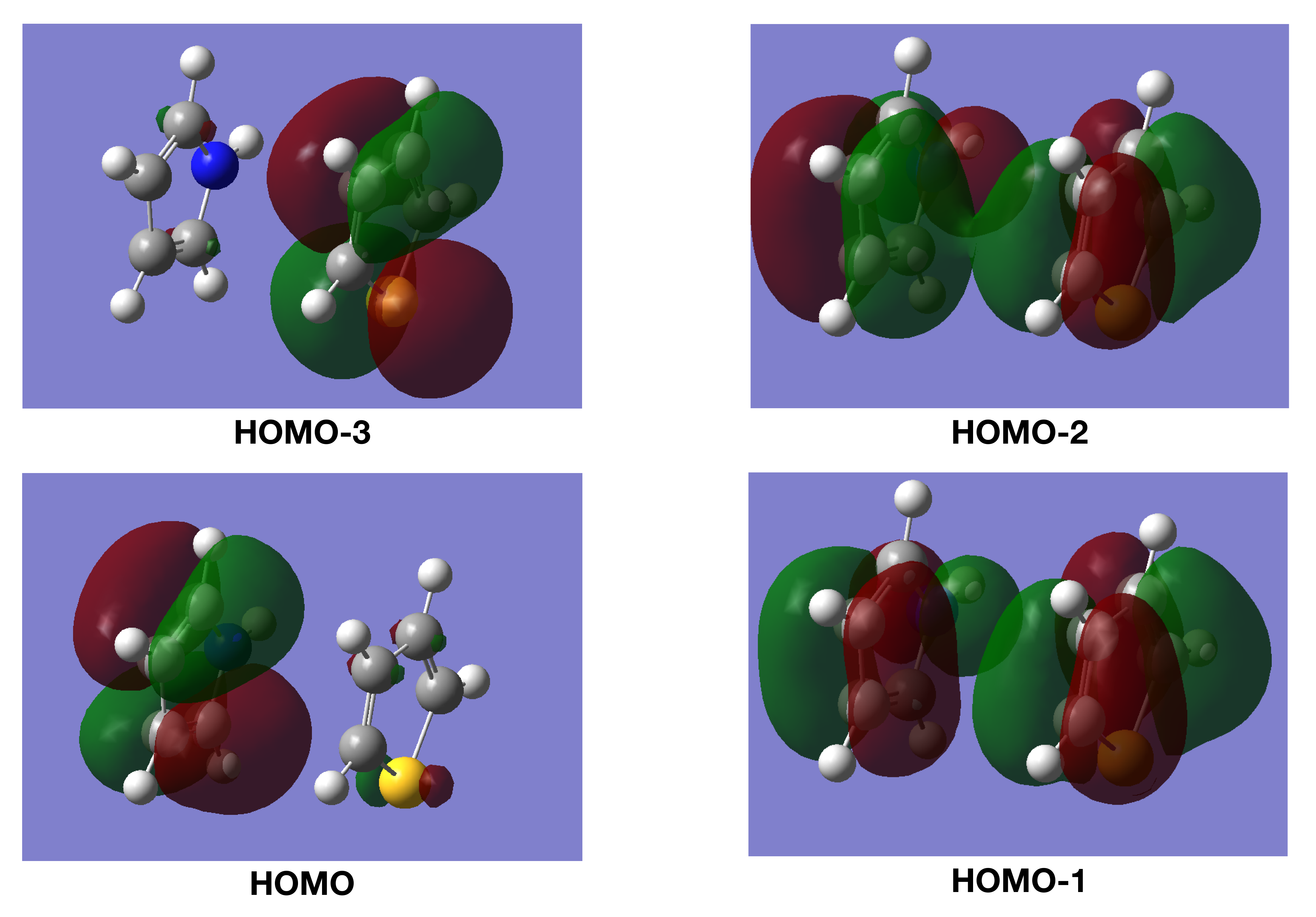}
\caption{HOMO-3 through the HOMO of a pyrrole-thiophene pair 
at 4.2 \AA\ separation and 90 degrees rotation.}
\label{fig:ThPyrOrbitals90}
\end{center}
\end{figure}

Fig.\ \ref{fig:ThPyrOrbitals90}
shows the HOMO-3 through HOMO
of the thiophene-pyrrole pair at 4.2 \AA\ and 90 degrees rotation.
This rotation is optimally aligned
for the pyrrole HOMO-1 and thiophene HOMO to mix,
forming the bonding and antibonding HOMO-2 and HOMO-1 pair orbitals.
At 90 degrees rotation,
the HOMO of thiophene has its lobes directly across from the HOMO-1 of pyrrole. 

\begin{figure}
     \centering
     \begin{subfigure}[b]{0.4\textwidth}
         \centering
         \includegraphics[width=\textwidth]{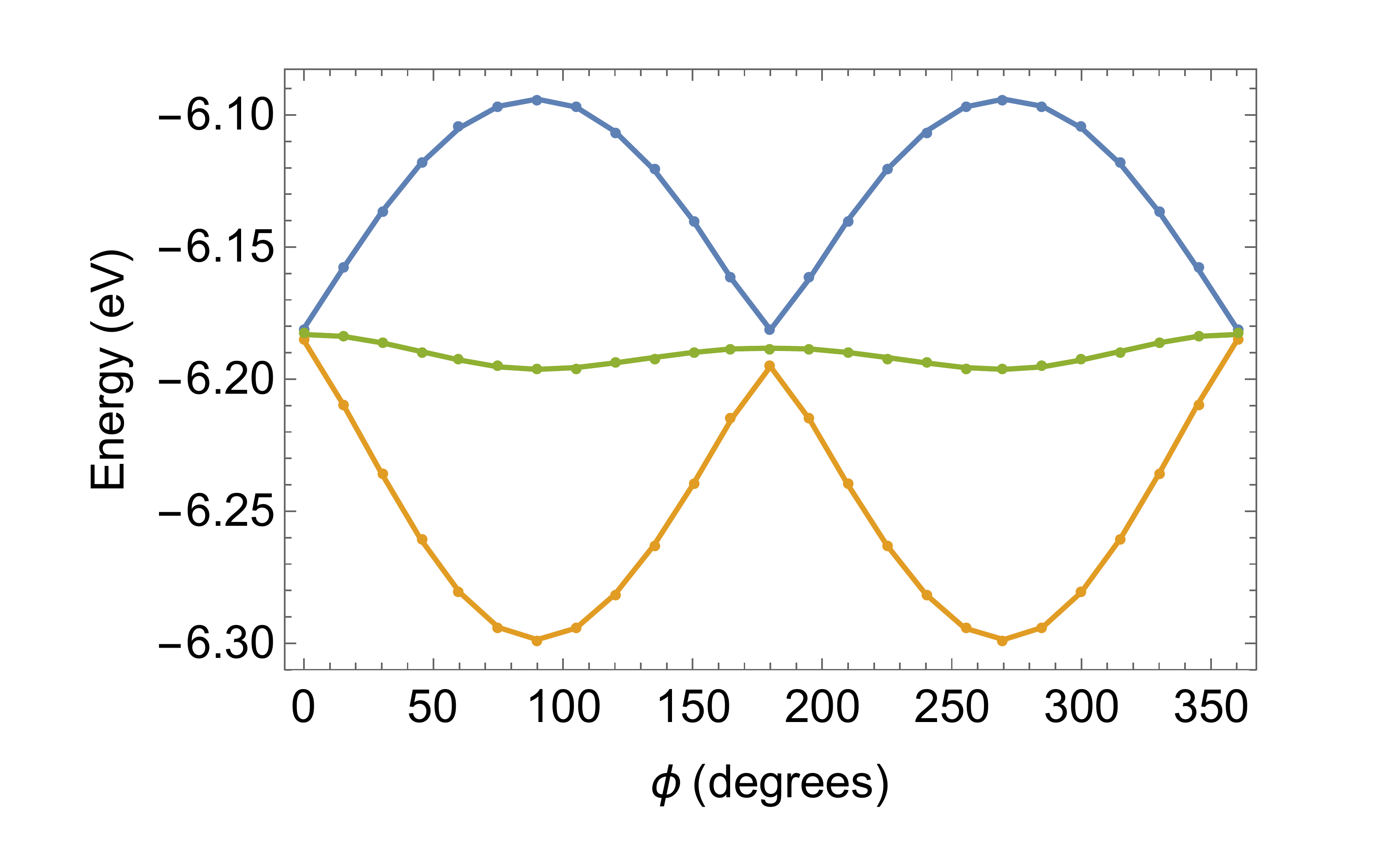}
         \caption{}
         \label{fig:ThPyrRotEnerg}
     \end{subfigure}
     \hfill
     \begin{subfigure}[b]{0.4\textwidth}
         \centering
         \includegraphics[width=\textwidth]{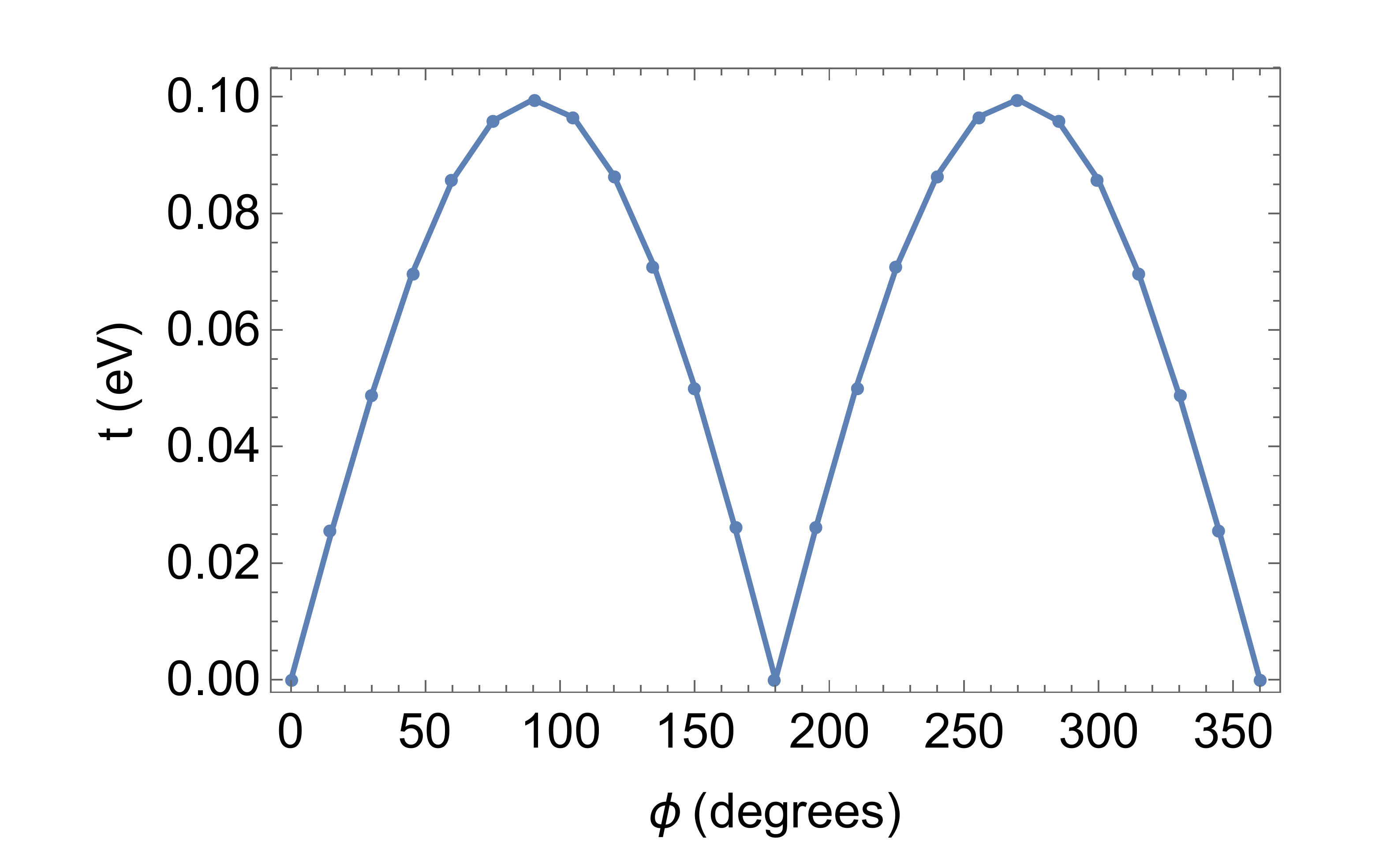}
         \caption{}
         \label{fig:ThPyrRott}
     \end{subfigure}
	\caption{(a): Pair HOMO-1 (blue) and HOMO-2 (orange) energies for thiophene and pyrrole scan 
	at 4.2 \AA\ separation and their average (green) versus $\phi$. (b): Resulting $t$ value versus $\phi$.}
	\label{fig:ThPyrRotScan}
\end{figure}

Fig.\ \ref{fig:ThPyrRotScan} presents results 
for the pair HOMO-2 and HOMO-1 energies,
from which the $t$ to hop from the pyrrole HOMO-1 to the thiophene HOMO 
is computed using our new method. 

\section{Thiophene tilt scans}

We now return to the thiophene-thiophene pair 
to explore the effect on $t$ of ``tilting'' one monomer,
i.e., rotating it about an axis other than the common normal direction
that connects the centers of mass of the two molecules in their initial configuration.
Although the monomers are identical, the ESD method fails for this scan 
because the ``tilt'' rotation breaks the C2 symmetry of the configuration;
thus, the Frontier Orbital Numerical Projection method will be used to calculate $t$. 

\begin{figure*}
\begin{center}
\includegraphics[width=0.9\textwidth]{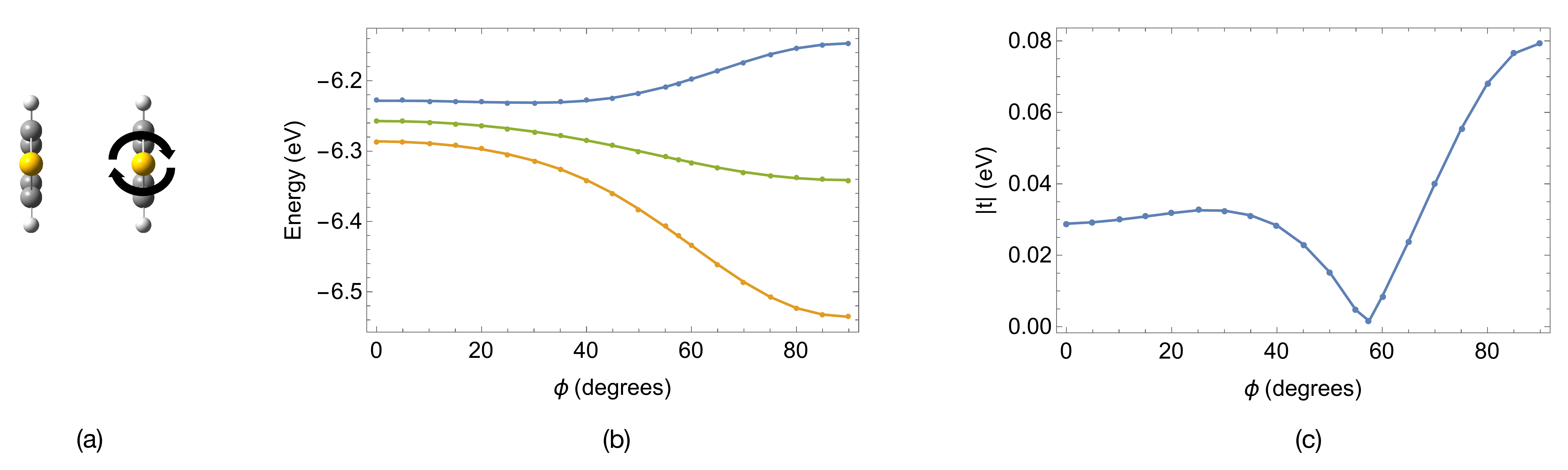}
\caption{(a): Schematic of tilt scan 1, at 5.0 \AA\ separation. 
(b): Pair HOMO (blue) and HOMO-1 (orange) energies and their average (green) versus $\phi$. 
(c): Resulting $t$ value versus $\phi$.}
\label{fig:ThThTiltScan1}
\end{center}
\end{figure*}

The first tilt scan rotates one of the two thiophenes
around the axis from the center of the ring to the sulfur,
as shown in Fig.\ \ref{fig:ThThTiltScan1}(a).
Fig.\ \ref{fig:ThThTiltScan1} shows the energy of the pair HOMO and HOMO-1,
and the corresponding value of $|t|$ throughout the tilt.
For this scan, $t$ changes sign as the rotation angle $\phi$ exceeds 57.5 degrees.
This reflects a change in the relative sign 
of the lobes of the two monomer HOMOs that are in close proximity, described further below.
The results presented in Fig.\ \ref{fig:ThThTiltScan1} are for the absolute value of $t$,
which is why the graph of Fig.\ \ref{fig:ThThTiltScan1}(c) 
decreases linearly to zero at around 57.5 degrees, 
and abruptly rises again.

As the molecule tilts, lobes that were positioned to hybridize
move away from their initial optimal position at $\phi=0$. 
At about 57.5 degrees, the lobes on the tilted thiophene 
are sandwiched between the lobes on the other thophene,
such that the pair is neither bonding nor antibonding.
Beyond 57.5 degrees, the lobes line up to favor hybridization in a new arrangement.
The sides of the two closest lobes on the rotated molecule
begin to align across from the fronts of the two closest lobes on the stagnant molecule,
again resulting in bonding and antibonding orbitals.

\begin{figure}[htbp]
\begin{center}
\includegraphics[width=0.4\textwidth]{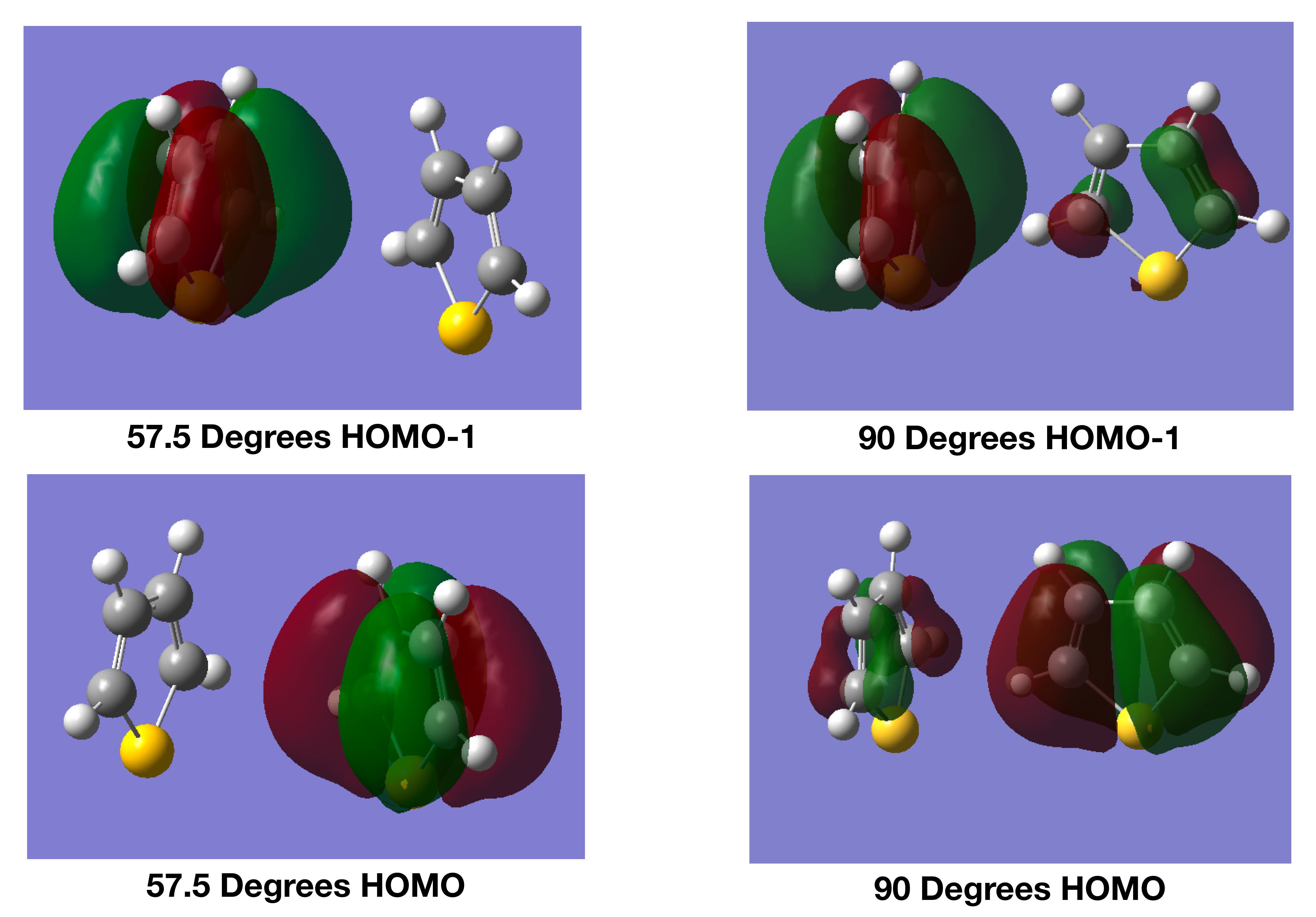}
\caption{HOMO-1 and HOMO of thiophene-thiophene pair of molecules 
with 57.5 and 90 degrees tilt and 5.0 \AA\ separation.}
\label{fig:ThThTilt1Orbitals}
\end{center}
\end{figure}

Fig.\ \ref{fig:ThThTilt1Orbitals} demonstrates both the negligible hybridization at 57.5 degrees
and the hybridization between the sides of the lobes of one molecule and the fronts of the other
that begins at rotations past 57.5 degrees.

\begin{figure*}
\begin{center}
\includegraphics[width=0.9\textwidth]{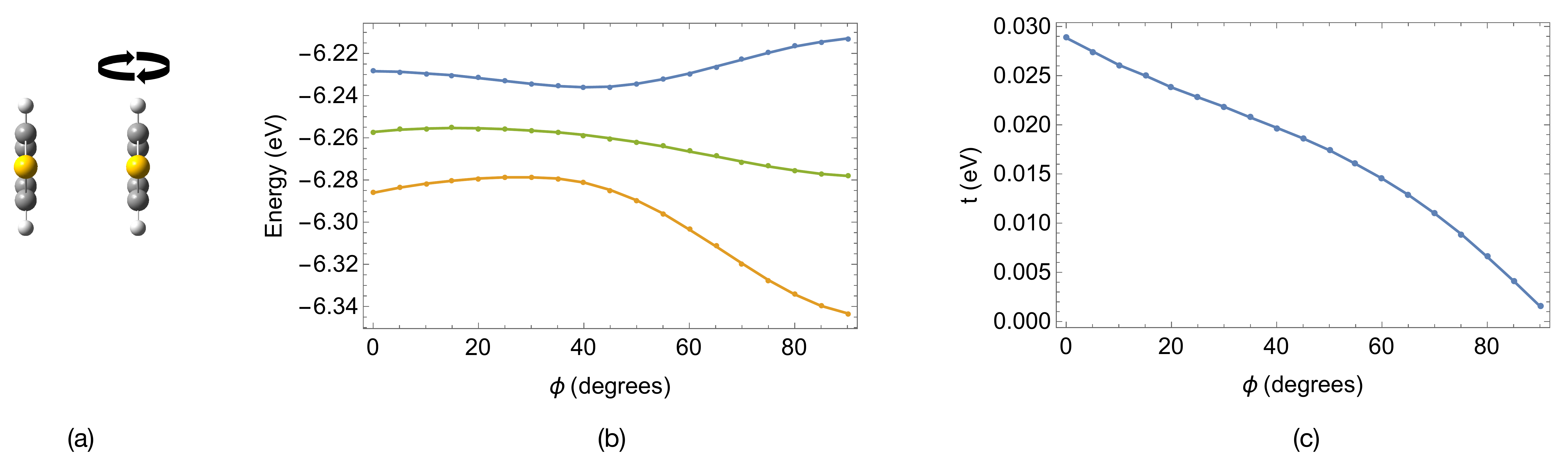}
\caption{(a): Schematic of thiophene tilt scan 2 at 5.0 \AA\ separation. 
(b): Pair HOMO (blue) and HOMO-1 (orange) energies and their average (green) versus $\phi$. 
(c): Resulting $t$ value versus $\phi$.}
\label{fig:ThThTiltScan2}
\end{center}
\end{figure*}

\begin{figure}[htbp]
\begin{center}
\includegraphics[width=0.4\textwidth]{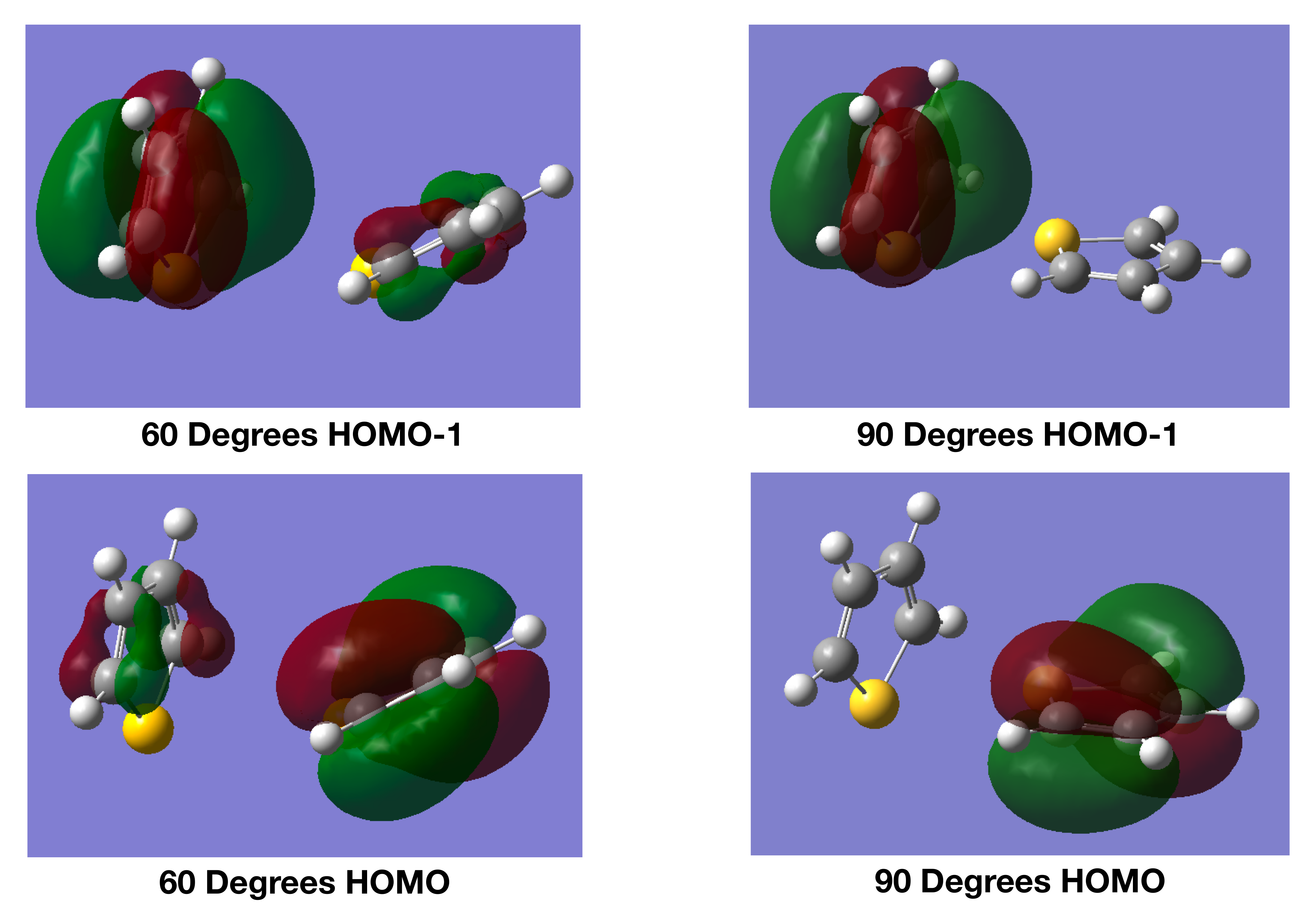}
\caption{HOMO-1 and HOMO of thiophene-thiophene pair of molecules 
with 60 and 90 degrees tilt and 5.0 \AA\ separation.}
\label{fig:ThThTilt2Orbitals}
\end{center}
\end{figure}

We can also tilt one thiophene about the axis in the plane of the ring
orthogonal to the previous tilt axis,
which tips its sulfur to point towards the other thiophene,
as shown in Fig.\ \ref{fig:ThThTiltScan2}(a).
$t$ now decreases throughout the entire scan (see Fig.\ \ref{fig:ThThTiltScan2}(c)).
Fig.\ \ref{fig:ThThTilt2Orbitals} shows that for this tilt scan,
the monomer orbitals never realign to a geometry consistent with hybridization.

\section{Transferability}

Evidently, interactions between large molecules should not be analyzed
as if they could be projected down to a single frontier orbital on each.  
Large molecules will have multiple low-lying orbitals which must be represented.  
For example, all-trans thiophene oligomers will have a set of close-lying sinusoidal orbitals, 
which in the limit of long oligomers becomes a 1d band.

As we have emphasized, our method is meant to be used in conjunction with tight-binding models.  
In such models, one makes appropriate choices of ``sites'' as moieties of a reasonable size, 
such that representing them by a single localized orbital is a good approximation.  
Properly constructed tight-binding models exhibit good agreement with DFT calculations,
and reasonable transferability of their coefficients 
(onsite energies and hopping matrix elements) 
between molecules of different architecture built from the same moieties. 
\cite{Tipirneni:2020hg}.

Therefore, an important application of our method 
to compute hopping matrix elements between nearby moieties
is to supply coefficients for tight-binding models of larger molecules.
An important question that arises in such applications
is the extent to which hopping matrix elements are ``transferable'',
i.e., can be applied between the same two local elements 
in different molecular structures.

\begin{figure}[htbp]
\begin{center}
\includegraphics[width=0.3\textwidth]{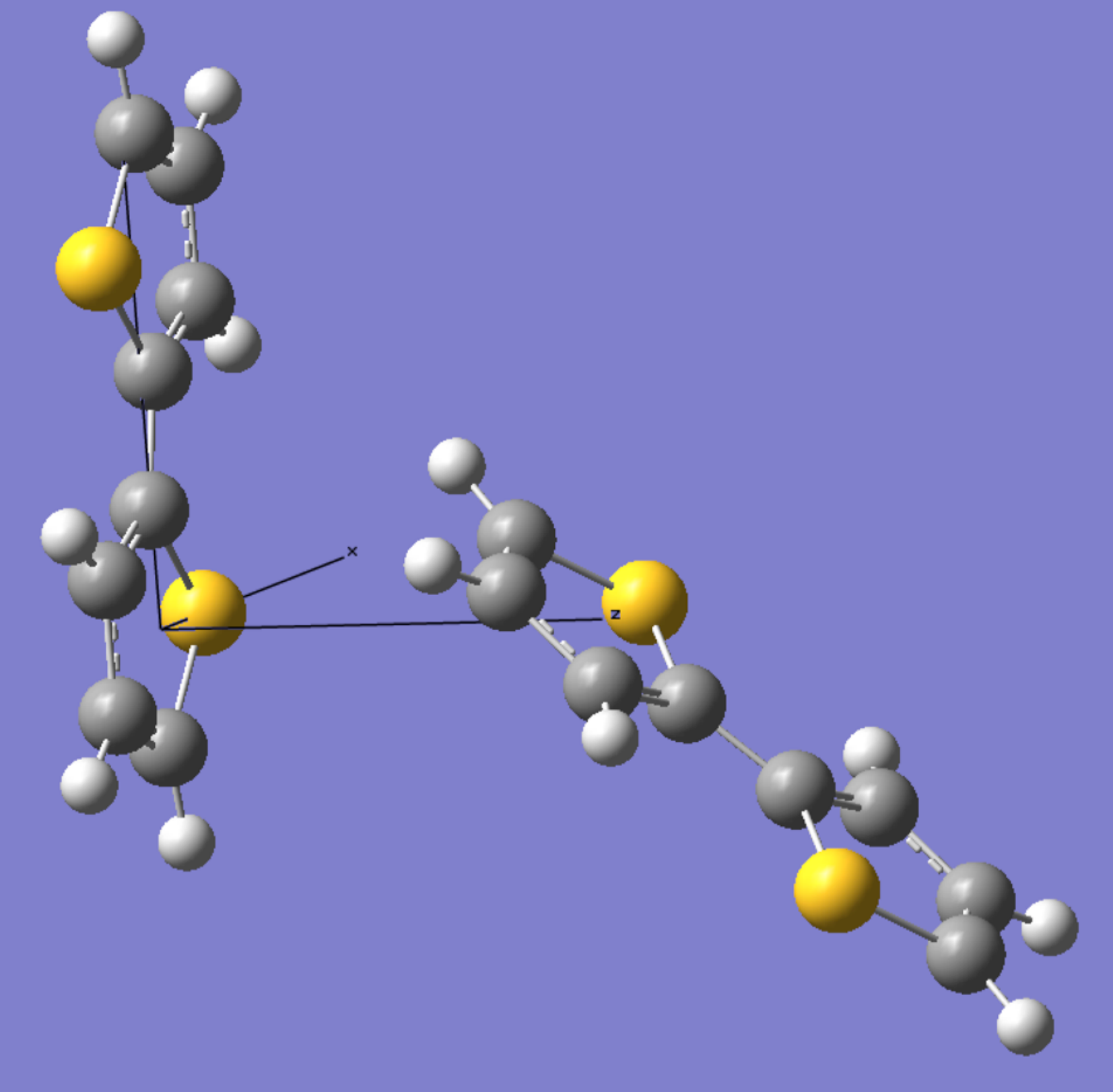}
\caption{Two thiophene dimers, with one tilted with respect to the other by 45 degrees,
analogous to ``tilt scan 1'' of Fig.\ \ref{fig:ThThTiltScan1} for thiophene monomers.}
\label{fig:dimerTiltGeom}
\end{center}
\end{figure}

To provide a simple test of transferability, 
we perform ``tilt scan 1'' of Fig.\ \ref{fig:ThThTiltScan1} again, 
but this time using thiophene dimers rather than monomers.
In the scan, the end monomers of each dimer 
are positioned and rotated in the same way as before;
Fig.\ \ref{fig:dimerTiltGeom} shows an example of the resulting geometry,
for a tilt angle of 45 degrees.

\begin{figure}[htbp]
\begin{center}
\includegraphics[width=0.5\textwidth]{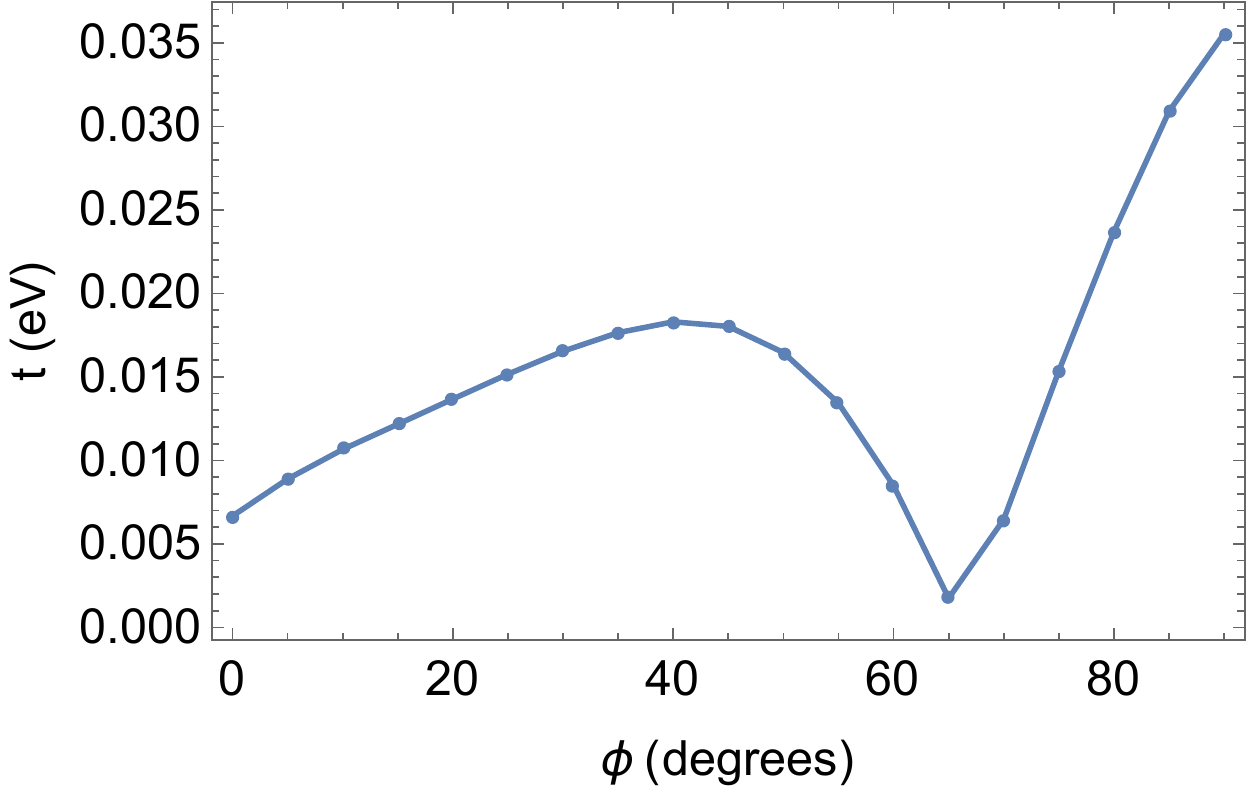}
\caption{Hopping matrix element $|t|$ versus tilt angle,
for a pair of thiophene dimers, one tilted with respect to the other,
in the same way as for the tilt scan results of Fig.\ \ref{fig:ThThTiltScan1}.}
\label{fig:dimerTiltResults}
\end{center}
\end{figure}

On physical grounds, we expect that hopping between the molecules
occurs between the two end monomers,
so that the hopping matrix element obtained from this scan
should be somehow related to the results of Fig.\ \ref{fig:ThThTiltScan1}.
Fig.\ \ref{fig:dimerTiltResults} shows $|t|$ versus tilt angle for the new scan;
the curve very much resembles the result of Fig.\ \ref{fig:ThThTiltScan1},
but weaker by about a factor of two.
As we will now show, that is precisely what we expect 
based on a simple tight-binding model.

The pair of interacting dimer molecules 
can be described by a tight-binding Hamiltonian of the form
\begin{equation}
H^{(4)} = \left(\begin{array}{cccc}
\epsilon_{1T} & -t & 0 & 0 \\
-t & \epsilon_{2T} & 0 & -t' \\
0 & 0 & \epsilon_{1B} & -t \\
0 & -t' & -t & \epsilon_{2B}
\end{array}\right)
\label{eq:intDimer}
\end{equation}
In writing Eqn.\ \ref{eq:intDimer}, we allow for the possibility
that the onside energies of sites 1 and 2 
on the top (T) and bottom (B) two molecules
may be shifted from their values in isolation
by the proximity of the other molecule.
$t'$ is the hopping matrix element 
between the two nearby moieties (site 2 on T and B);
the intramolecular hopping matrix elements $t$
are assumed unaffected and identical.

We can relate the 4-site Hamiltonian $H^{(4)}$
to a simpler one that describes only interactions
between the HOMO frontier orbitals of the two molecules,
by projecting onto the $2 \times 2$ subspace 
spanned by the two isolated HOMO orbitals:
\begin{equation}
\psi_T = \left(\begin{array}{c} 1/\sqrt{2} \\ -1/\sqrt{2} \\ 0 \\ 0 \end{array}\right) , \quad 
\psi_B = \left(\begin{array}{c} 0 \\ 0 \\ 1/\sqrt{2} \\ -1/\sqrt{2} \end{array}\right) 
\end{equation}

The relevant matrix elements are 
\begin{eqnarray}
\langle \psi_T | H^{(4)} | \psi_T \rangle &=& (1/2)(\epsilon_{1T} + \epsilon_{2T}) + t \equiv \bar \epsilon_T \nonumber \\
\langle \psi_B | H^{(4)} | \psi_B \rangle &=& (1/2)(\epsilon_{1B} + \epsilon_{2B}) + t \equiv \bar \epsilon_B \nonumber \\
\langle \psi_T | H^{(4)} | \psi_B \rangle &=& -t'/2
\end{eqnarray}
Thus the projected 2-site Hamiltonian $H^{(2)}$ takes the form
\begin{equation}
H^{(2)} = \left(\begin{array}{cc} \bar \epsilon_T & -t'/2 \\ -t'/2 & \bar \epsilon_B \end{array}\right)
\end{equation}
This is exactly what we wrote for analyzing the hopping between two nearby moieties,
 with the hopping matrix element is identified as $t'/2$, 
 half of the local value acting between site 2 on the top and bottom molecules.
 So by applying the Frontier Orbital Numerical Projection method
 to the pair of dimer molecules in a tilt scan,
 we should expect to obtain a hopping matrix element 
 that couples the HOMOs of the two molecules,
 with a value about half that obtained 
 in the corresponding scan for a pair of thiophene monomers.

\section{Conclusions}

Computing hopping matrix elements for moieties in close proximity
is an important part of building tractable models
for electronic interactions between such moieties,
with applications in modeling carrier transport
as well as exciton energetics and polarization
in organic semiconductor materials.
Hopping matrix elements can be inferred 
from the energy level splittings 
of hybridized frontier orbitals for the interacting pair
only in very special cases,
for which the molecules are chemically identical
and symmetrically placed,
leading to bonding and antibonding orbitals
that are symmetric and antisymmetric combinations
of the constituent monomer orbitals.

To treat the vast preponderance of cases
in which this symmetry is not present,
a more versatile method is required.
In this work, we present a simple and convenient method
to compute hopping matrix elements 
between any pair of nearby small molecules.
The method relies on computing overlap integrals
between the frontier orbitals of the pair
and the frontier orbitals of the two interacting molecules
that hybridize to form the pair orbitals.

Such overlap integrals can in principle be computed analytically
from the set of basis function expansion coefficients for the orbitals.
However, commonly available DFT platforms such as Gaussian
do not provide utilities for such calculations;
code to compute such overlaps is tedious to write and validate,
and not readily available.
Instead, we compute such overlap integrals numerically using ``cube files'', 
or spatially discretized representations of the pair and monomer orbitals.
Such cube files can be readily obtained 
from utilities supplied with Gaussian and other DFT platforms.
Following this approach, we have created a set of Python and bash shell scripts
to compute hopping matrix elements between nearby molecules,
and to carry out scans in which one constituent molecule
is progressively translated or rotated with respect to the other
(see Supplemental Information for details).

We have validated this Frontier Orbital Numerical Projection method
by repeating published results for scans of $t$ versus separation
between two parallel tetracene molecules
(for which energy splitting is sufficient to determine $t$),
and as well for $t$ versus rotation between two initially parallel ethylene molecules
(for which orbital overlaps must be computed to determine $t$);
we find good agreement with previous results.
Then, we demonstrate how the method can be used
to explore how hopping matrix elements vary
with the chemical identity and relative placement
of two molecules in close proximity.
We select for our examples moieties that occur in the structure 
of organic semiconducting polymers and non-fullerene acceptors
currently under study for photovoltaic material applications.

In forthcoming work, we apply our method to explore 
the dependence of hopping matrix elements on relative placement of moieties 
extracted from large configurations of semiconducting polymers
generated by atomistic molecular dynamics simulations.
Such studies will give insight into the heterogeniety 
of electronic coupling within these disordered molecular solids,
and help us identify those configurations that lead
to particularly large or small values of $t$.

\section{Acknowledgements}

Funding support from the Office of Naval Research through Award N00014-19-1-2453 is gratefully acknowledged.

\section{Appendix:  non-orthogonal basis}

In our simple analysis of frontier orbitals on a pair of nearby moieties
interacting through a hopping matrix element,
we have implicitly assumed the orbitals $ \psi_1$ and $\psi_2 $
are far enough apart that they may be regarded as orthogonal.
If they are, then the expansion coefficients of a dimer frontier orbital $\psi $,
written as 
\begin{equation}
\psi  = c_1 \psi_1+ c_2 \psi_2 
\end{equation}
can be computed straightforwardly as 
\begin{equation}
c_i = \langle \psi_i | \psi \rangle
\end{equation}
As emphasized by Valeev et al., this approach should be modified 
if the localized states on the two moieties 
have significant direct overlap\cite{Valeev2006}.

If $\psi_1$ and $\psi_2$ are not quite orthogonal, 
we can still compute $c_1$ and $c_2$,
by introducing a ``dual basis'' of wavefunctions 
$\tilde \psi_1$ and $\tilde \psi_2$, with the properties
\begin{equation}
\langle \tilde \psi_i | \psi_j \rangle = \delta_{ij}
\end{equation}
We can use the dual wavefunctions $\tilde \psi_i$
to compute the expansion coefficients $c_i$ as
\begin{equation}
c_i = \langle \tilde \psi_i | \psi \rangle
\end{equation}

To compute the dual basis in terms of the overlap integral 
$\langle \psi_1 | \psi_2 \rangle = \gamma$,
we expand the dual wavefunctions in terms of the original set as
\begin{eqnarray}
\tilde \psi_1 &=& \alpha_1 \psi_1 + \beta_1 \psi_2 \nonumber \\
\tilde \psi_2 &=& \beta_2 \psi_2 + \alpha_2 \psi_2
\end{eqnarray}
The dual basis conditions lead to a matrix equation 
for the coefficients $\alpha_i$ and $\beta_i$:
\begin{equation}
\left(\begin{array}{cc}\alpha_1 & \beta_1 \\ \beta_2 & \alpha_2 \end{array}\right)
\left(\begin{array}{cc}1 & \gamma \\ \gamma & 1 \end{array}\right)
= \left(\begin{array}{cc} 1 & 0 \\ 0 & 1 \end{array}\right)
\end{equation}
That is, the matrix of expansion coefficients
is the inverse of the matrix of overlap integrals $\langle \psi_i | \psi_j \rangle$.

Evidently for the present $2 \times 2$ case,
we can take $\beta_1 = \beta_2 = \beta$, and $\alpha_1 = \alpha_2 = \alpha$.
Explicitly computing the inverse, we find
\begin{eqnarray}
\alpha &=& 1/(1-\gamma^2) \nonumber \\
\beta &=& -\gamma/(1-\gamma^2)
\end{eqnarray}

If $\psi_1$ and $\psi_2$ are nearly orthogonal, $\gamma$ is small,
and we have $\alpha \approx 1$, $\beta \approx 0$ 
so that $\tilde \psi_1 \approx \psi_1$.
More generally, the expansion coefficients can be given as
\begin{eqnarray}
c_1 &=& \frac{\langle \psi_1 | \psi \rangle - \gamma \langle \psi_2 | \psi \rangle}{1 - \gamma^2} \nonumber \\
c_2 &=& \frac{\langle \psi_2 | \psi \rangle - \gamma \langle \psi_1 | \psi \rangle}{1 - \gamma^2}
\end{eqnarray}

Because we consider only pairs of moieties that are not too close to each other,
the overlap integrals in all the scans we report here are quite small.
For example, in the ``tilt scan 1'' leading to the results of Fig.\ \ref{fig:ThThTiltScan1},
the value of $\gamma$ is everywhere less than 0.005 in magnitude.
Accordingly, we neglect the corresponding small correction in our reported results.

\bibliography{hoppingBib}
\end{document}